\documentclass[prb,twocolumn,showpacs,amsmath,amssymb,superscriptaddress,preprintnumbers,floatfix]{revtex4}

\usepackage{graphicx,psfrag}
\newcommand{\bA}{{\bf A}}
\newcommand{\ba}{{\bf a}}
\newcommand{\bG}{{\bf G}}
\newcommand{\bk}{{\bf k}}
\newcommand{\bl}{{\bf l}}

\newcommand{\bp}{{\bf p}}
\newcommand{\bpi}{{\boldsymbol{\pi}}}
\newcommand{\bq}{{\bf q}}
\newcommand{\bQ}{{\bf Q}}
\newcommand{\bR}{{\bf R}}
\newcommand{\br}{{\bf r}}
\newcommand{\bv}{{\bf v}}

\newcommand{\bZ}{{\bf Z}}
\newcommand{\btau}{{\boldsymbol{\tau}}}
\newcommand{\dphi}{\delta\phi}

\newcommand{\Psibar}{{\overline{\Psi}}}

\newcommand{\hatV}{\hat{V}}
\newcommand{\cA}{{\cal A}}
\newcommand{\eps}{\epsilon}

\begin{document}

\title{Dirac-Bogoliubov-deGennes quasiparticles in a vortex lattice}

\author{Ashot Melikyan}
\email{ashot@phys.ufl.edu}
\affiliation{Institute of Fundamental Theory, Department of Physics, University of Florida, Gainesville, FL, 32611}
\author{Zlatko Te\v{s}anovi\'{c}}
\email{zbt@pha.jhu.edu}
\affiliation{Department of Physics and Astronomy, Johns Hopkins University, Baltimore, MD, 21218}

\date{\today}

\begin{abstract}
In the mixed state of an extreme type-II  $d$-wave superconductor and
within a broad regime of weak magnetic fields ($H_{c1}\ll H\ll H_{c2}$), the 
low energy Bogoliubov-deGennes quasiparticles can be effectively
described as Dirac fermions moving in the field of singular scalar and
vector potentials. Although the effective linearized
Hamiltonian operator formally
does not depend on the structure of vortex cores, a singular nature of the
perturbation requires choosing a self-adjoint extension of the Hamiltonian by
imposing additional boundary conditions at
vortex locations. Each vortex is described by a single parameter $\theta$
that effectively represents all effects arising from the physics
beyond linearization. With the value of $\theta$ properly fixed,
the resulting density of states of Dirac Hamiltonian
exhibits full invariance under
arbitrary singular gauge transformations applied at vortex positions.
We identify the self-adjoint extensions of the
solutions found earlier, within the framework of the linearized Hamiltonian
diagonalized by expansion in the plane wave basis, and
analyze the relation between 
fully self-consistent formulation of the problem and the linearized model. 
In particular, we construct the low-field scaling form of the 
nodal quasiparticle
spectra which incorporates the self-adjoint extension parameter $\theta$
explicitly and generalizes the conventional Simon-Lee scaling. 
In a companion paper, we also present a detailed numerical study of the 
lattice $d$-wave superconductor model
and examine its low energy, low magnetic field behavior with an eye on
determining the proper self-adjoint extension(s) of the
linearized continuum limit. In general, we find that the density of
quasiparticle states always vanishes at the chemical potential, either
linearly or by virtue of a finite gap. The low energy continuum limit is thus
faithfully represented by Dirac-like fermions which are either 
truly massless, massless at the linearized level 
(mass $\sim H$) or massive (mass $\sim\sqrt{H}$), depending on
the mutual commensuration of magnetic length and lattice spacing.


\end{abstract}


\maketitle

\section{Introduction}
In conventional $s$-wave superconductors the BCS quasiparticle 
excitation spectrum is 
gapped and, in presence of a vortex defect, the 
quasiparticles form a discrete set of  
Caroli-deGennes-Matricon states\cite{caroli}. These states are
bound below the gapped continuum and are described by 
wavefunctions that decrease exponentially away from the vortex position.
Once finite density of vortices is induced by a magnetic field, as is the
case in the mixed state of type-II superconductors, these discrete levels
broaden into extremely narrow quasiparticle bands whose effect on
bulk properties is rather modest, as long as the magnetic field $H$
remains well below the upper critical field $H_{c2}$. 
The situation is entirely different in high temperature
cuprate superconductors: there
the quasiparticle gap has a $d_{x^2-y^2}$ symmetry
and vanishes at four nodal points on the Fermi surface. The excitation
spectrum is gapless, with linearly vanishing density of states at the
Fermi level. As a consequence, 
there are no strictly bound states near a vortex defect in such a
superconductor\cite{ft98}. Instead, the quasiparticle wavefunctions
exhibit a power law decrease away from vortex position and the spectrum
is continuous\cite{footi}. Once
a finite magnetic field is turned on and vortex lattice appears in the
mixed state, the low energy quasiparticles are expected to form broad
bands which dramatically influence thermodynamics, transport
and vortex dynamics of high temperature superconductors.

After an early semiclassical approximation 
due to Volovik\cite{volovik}, who
suggested that quasiparticles spectrum undergoes 
a Doppler shift
$E_{\bk}\to E_{\bk}-\bv (\br)\cdot\bk$ due to
the vortex-induced superflow $\bv (\br)$ at a given
point $\br$, the quantum theory of quasiparticles in the presence of vortex
lattice evolved along two somewhat separate lines. On one side,
the self-consistent Bogoliubov-deGennes (BdG) equations for $d$-wave
superconductor were solved numerically in
continuum, both for a single vortex\cite{ft98} and for a vortex
lattice\cite{yasui_kita}. Additionally, the numerical solution
was also obtained for BdG equations of a tight-binding
lattice $d$-wave superconductor with boundary conditions corresponding to
a periodic vortex array\cite{wang_macdonald}.
In all cases it was clearly demonstrated that a single vortex is
incapable of truly localizing $d$-wave quasiparticles and broad low energy 
bands are evident in the quasiparticle excitation spectrum of a vortex
lattice.

Solving BdG equations in the inhomogeneous
mixed state, particularly for a $d$-wave
superconductor with its extended low energy quasiparticle states,
is a daunting task. It is natural to inquire whether a simpler, analytic
description might be devised which will allow one to address important
and realistic physics questions including disordered vortex lattice, thermal
and quantum fluctuating vortices, etc., where numerical solutions are
either too forbidding or too opaque to be useful. An important step
along this path is the linearized version of the BdG 
Hamiltonian introduced by Simon and Lee\cite{simonlee}. Since the
homogeneous system at zero field can be described by four species 
of massless two-component Dirac fermions residing at nodal points
on the Fermi surface, it was argued that in low magnetic fields
$H_{c1}\ll H\ll H_{c2}$, when
the separation between vortices given by magnetic length $l$ 
becomes much larger
than the size of the vortex cores -- set by
the BCS coherence length $\xi$ -- the properties of low
energy  quasiparticles to the leading order in $\xi/l$ can still be
described by a superposition of 
four independent Hamiltonians at each node\cite{simonlee,ft00}.

By devising an eponymous gauge transformation, 
Franz and Te\v sanovi\' c\cite{ft00} recast the
nodal BdG Hamiltonian as that of 
a Dirac particle moving in a zero {\em average} magnetic field and
subject to effective long-range scalar and vector potentials whose 
point-like sources are located at vortex positions. This approach
revealed that nodal quasiparticles couple to
vortices through a {\em combined} effect of {\em two} long-range terms:
the semiclassical Doppler shift must
be accompanied by a purely quantum
mechanical Berry phase, which attaches a $\exp(\pm i\pi)=(-1)$ phase
factor to a wavefunction of a quasiparticle encircling
an $hc/2e$ vortex\cite{ft00,vmft}.  The FT transformation
has been widely used to directly address, by combination
of general symmetry arguments\cite{marinelli,vmft,ashvinii,ftv}
and explicit analytic and numerical 
calculations,
\cite{marinelli,vmft,ashvini,knapp,goryo,ftv,herbut,ye,kkk,lages,lagesi,nikolic}
various aspects of quantum mechanics of nodal
quasiparticles in presence of vortices and will serve as the
departure point for our discussion\cite{footii}.
We should stress, however, 
that the results and conclusions of this paper
are completely general and remain unaffected if instead
of the FT transformed BdG Hamiltonian one uses more familiar
framework of the magnetic translations group applied to the
original BdG equations.
A natural question is why should not one 
simply use the magnetic translation group (MTG) and forgo the FT transformed 
problem altogether? The answer is twofold: 
obviously, in the FT transformed case
one is dealing with representations of ordinary Bloch translation group
that are far more convenient than those of MTG. More importantly,
the FT transformation is custom tailored for an efficient extraction 
of the low-energy, long-distance sector of the original problem,
a prime feature in numerous physical situations. 

While the problem of $d$-wave nodal quasiparticles interacting with vortex
defects is of considerable theoretical interest in itself, its rich
phenomenology and relevance for numerous experiments on cuprates
promote it to the forefront of modern condensed matter 
physics. In high temperature superconductors, the low field
regime ($H_{c1}\ll H\ll H_{c2}$) of the vortex state covers
a large portion of the $H-T$ phase diagram and a large body
of experimental work thought to offer clues regarding microscopic
origins of high temperature superconductivity 
is available both for analysis and future
refinements. In particular,
the high resolution STS/STM measurements of 
tunneling conductance in YBCO and BSCCO, which is
proportional to the quasiparticle local density of states (LDOS), 
reveal peaks near vortices that appear
at energies $5.5$ meV in YBCO and $7$ meV in BSCCO and at magnetic fields 
$\sim 6$ Tesla\cite{fischer}.
The origin of the LDOS peaks is still a hotly debated subject and several
scenarios\cite{arovas97,ft98,andersen,ft01,han,demler,wang,zhu_ting01,berthod,sheehy,nayak,nguyen,bishop,kishine,balatsky,ioffe,ghosal,ogata}
describing various microscopic physics inside and 
around vortex cores, such as 
spin- and charge- density waves (SDW and CDW), Mott insulator,
$d$-density wave and others, have been 
proposed  to explain these features. This structure in the quasiparticle
LDOS appears related to the experiments 
on neutron scattering\cite{lake} which
hint at the development of an antiferromagnetic
order parameter localized in the vicinity of vortex cores. A 
mean-field factorization of $t-J$ model on a tight-binding
lattice and allowing for both $d$-wave superconductor and AF orders
does indeed, for a certain range of parameters in the self-consistent
solution, describe a $d$-wave vortex state with an AF order
developing inside vortex 
cores\cite{arovas97,andersen,zhu_ting01,bishop,ghosal}. 
The details of the calculations, however, 
depend strongly on the parameters of the
model and the actual paring terms 
included in self-consistent approximation.
In fact it is not even clear that 
the interior of vortices can be reliably described within the
mean-field framework since the suppression of superconductivity inside
vortices is likely to lead to a formation of a strongly correlated
state characteristic of the pseudogap region in the phase diagram
of cuprates\cite{ft01,ioffe,ogata}.

The linearized continuum formulation of the bulk
properties of nodal quasiparticles has the distinct advantage 
of being essentially independent of such details pertaining to
the structure near vortex cores. Nodal BdG quasiparticle
behaves like a massless Dirac fermion and can be thought of
as ``critical''\cite{zirnbauer}: 
the Feynman path integral trajectories of such
particles lack any characteristic lengthscale. This ``criticality''
implies a certain degree of ``universality''\cite{durst,taillefer}:  
one expects that an effective low-energy theory of such Dirac fermions
can be constructed, which incorporates
all the  effects of intra-vortex physics on the bulk properties of the
BdG quasiparticles in terms of few simple parameters.
A detailed derivation of such an
effective theory, describing the influence of the physics in vicinity
of vortex cores on properties of nodal BdG quasiparticles in the bulk,
is the main goal of this paper.


Following FT, we start by representing 
the motion of nodal BdG quasiparticles in a vortex 
lattice by massless two-dimensional
Dirac fermions in combined vector and scalar potentials\cite{ft00}. 
Both vector and scalar potentials are
singular near vortices, and we will show that regularization of
this singular behavior is required, which itself depends
on the short-scale features of the physics near vortex cores. The
final effective theory is essentially the canonical FT Hamiltonian
supplemented by a boundary condition at each vortex. The variety of possible
physical scenarios describing different types of core and near-core
behaviors is represented by a {\em single}
dimensionless parameter $\theta\in[0,\pi)$ that characterizes 
the boundary condition and is associated with each vortex.
$\theta$ being dimensionless, such boundary conditions can in principle
modify the energy spectrum already at 
the leading order in $(k_Fl)^{-1}$  -- to
this order\cite{footiii} the value of $\theta$ uniquely determines the
full extent of the influence that the structure in or near vortex 
cores has on the bulk properties of nodal quasiparticles\cite{foottight}.
We generalize the conventional Simon-Lee scaling to 
include the explicit dependence on $\theta$
and numerically determine the scaling function.
This generalized scaling form for the linearized quasiparticle energy
levels is:
\begin{equation}
E_{n,{\bk}} = \frac{\hbar v_F}{l} 
{\cal E}_{n,{\bk}l}\left(\frac{v_F}{v_{\Delta}},\{\theta\}\right)~~,
\end{equation}
where  ${\cal E}$ is a universal dimensionless
function,  $n$  and $\bk$ denote the band number and the
Bloch momentum of the FT transformed states, respectively
 and $v_F/v_{\Delta}$ is the bare anisotropy of the 
$d$-wave nodes in a zero-field case.

We find three prominent behaviors of a 
$d$-wave superconductor as $H\to 0$ (while still $H\gg H_{c1}$)
and determine the associated self-adjoint extensions of the
linearized continuum problem:
the first is a gapless behavior with the Dirac nodes intact -- although
renormalized by the field -- but with their number {\em doubled} 
relative to the zero field result of four. This solution bears
considerable resemblance to the one reported 
earlier in Ref. \onlinecite{ft00}.
Second, we find new {\em gapped} spectra with no zero energy states. This
self-adjoint extension produces
a Dirac mass gap which scales as $1/l$ and is thus a
part of the leading order $H\to 0$ scaling function, unlike the gaps
of order $1/l^2$, which vanish in the leading order scaling and
are typically present even for the ``gapless'' case. The gapped
solution seems to be related to the ``interference'' gapped spectra
discussed in Refs. \onlinecite{vmft,tight}.
Finally, we find $\theta$ for which the quasiparticle bands
lack $(E,\bk)\to (-E,\bk)$ symmetry and for which the density of states
is finite at the Fermi level -- such solutions, however,
at least for the superconducting
gap parameter $\Delta$ not exceedingly small
in comparison to hopping $t$, as is appropriate for near optimally
doped or underdoped cuprates,
are never found in our numerical calculations on the lattice
$d$-wave superconductor. These tight-binding calculations,
reported separately\cite{tight}, always find either a linearly vanishing 
or gapped density of states at the Fermi level.
Thus, the so-called ``Volovik effect,'' a defining 
feature of an early semiclassical
approach, is effectively absent from a fully quantum mechanical solution.
It would be important to verify this feature of our results 
experimentally, for example, by precision measurements of a specific heat
or thermal conductivity.

Alternatively, the above regularization and associated self-adjoint
extensions can be implemented by introducing 
a fictitious potential which suppresses the wavefunctions of nodal
quasiparticles near vortices. This potential represents the nonlinear
corrections to the linearized BdG Hamiltonian. Because of the well-known
Klein paradox for Dirac particles, this potential cannot be chosen as a large
scalar potential barrier, and instead must be realized as a
position-dependent mass \cite{berry}, which 
grows rapidly as one approaches a vortex location but vanishes elsewhere. We
demonstrate that the two descriptions of the regularized FT Hamiltonians are
equivalent and yield essentially 
identical results in numerical calculations.
\psfrag{unitcell_A}{$A$}
\psfrag{unitcell_B}{$B$}
\psfrag{unitcell_a}{$(a)$}
\psfrag{unitcell_b}{$(b)$}
\psfrag{unitcell_eps}{$\eps$}
\begin{figure}[tbh]
\centering
\includegraphics[width=\columnwidth]{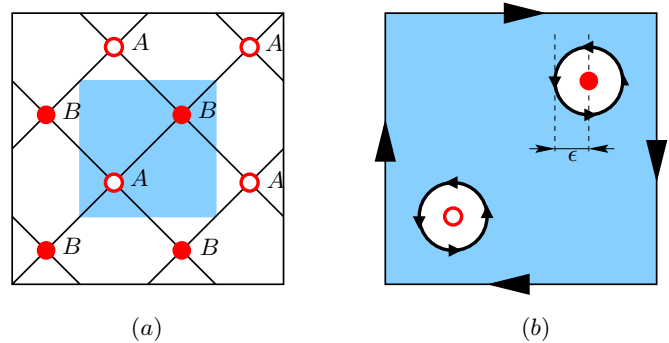}
\caption{\label{fig:unitcell} A magnetic unit cell containing two vortices.
Panel (b) shows contours used in the application of the Stokes theorem
discussed in section \ref{sae}.}
\end{figure}
The realization that the boundary conditions at
the vortices must be externally imposed resolves the difficulty
associated with previous attempts to 
find the spectrum of the quasiparticles numerically. 
Despite the gauge invariance of the 
linearized problem, the explicit form of the FT
Hamiltonian depends on the partition of the original vortex lattice into two
sublattices A and B. Such choice is clearly
not unique, and different selections
are related by singular gauge transformations. Obviously, all measurable
physical quantities, exemplified by the 
density of states (DOS), must be invariant under such
transformations. Surprisingly, numerical spectra of the
linearized FT Hamiltonian exhibit small but
persistent difference for distinct choices of A and B sublattices, 
when the wavefunctions are sought
as linear combinations of plane waves\cite{vmft} 
or when the Hamiltonian is solved by
discretization on a real space mesh\cite{marinelli,footiv}.
In what follows, we will demonstrate that the difficulties 
associated with the linearized model indeed are not
simply a  nuisance related  to the numerical procedure but rather
reflect a genuine dependence of the bulk properties of quasiparticles on the
internal structure of vortices. 
Even though the  bulk of 2D CuO planes is dominated by pure nodal $d$-wave
quasiparticles, the internal vortex structure affects the
spectrum through boundary conditions that should be imposed at the vortex
locations, due to singular nature of the
perturbation introduced by vortices, and the resulting energy spectra
for different vortex core types are generally distinct  
to the leading order in $\hbar v_F/l$.
We reanalyze the solutions found earlier, and find that they correspond to 
mutually different sets of boundary conditions, 
``spontaneously'' selected
by a numerical procedure used as well as
the choice of A and B sublattices. Once the fixed
boundary conditions are externally imposed, 
the numerical procedure that incorporates them 
generates results that are fully gauge independent.

The paper is organized as follows: in
the next two sections we introduce the linearized FT Hamiltonian
and discuss the necessity of imposing 
boundary conditions (BC) at vortex locations. 
We show how these BC are intimately related to the self-adjoint
extensions of the linearized BdG Hamiltonian\cite{vmft} and discuss such
extensions in detail in section \ref{sae}. After analyzing the
symmetry properties of the linearized problem with and  without
BC in section \ref{sym},  we describe in section \ref{num} a new
procedure used to address the problem, which incorporates the BC numerically.
Section \ref{relation}  establishes an alternative 
framework for regularization of
the linearized problem using mass potentials 
and discusses the connection between 
the full, non-linearized BdG equations and
self-adjoint extensions of the linearized model.
This is followed by a general discussion of the relation between
the $d_{xy}$ and $d_{x^2-y^2}$ continuum models in section \ref{orient}. 
Finally, we conclude with a brief summary of
our results.
\section{The Linearized Hamiltonian}
It was argued \cite{simonlee,ft00} that in the limit of weak magnetic
field $H\ll H_{c2}$ (but still $H\gg H_{c1}$),
the BdG Hamiltonian describing 
quasiparticles in an ideal vortex lattice can be
formally expanded\cite{footii} in powers 
of $(k_F l)^{-1}$ where magnetic length $l$ is
defined through the flux quantum $\Phi_0=hc/e$ as $l= \sqrt{\Phi_0/H}$.
In extreme type-II superconductors,  the regions separated from the 
vortices by distances sufficiently larger than vortex core size
$\sim \xi$  are accurately described by 
the center-of-mass superconducting order parameter
$\Delta(\br)=\Delta_0 e^{i\phi(\br)}$ with 
a nearly uniform amplitude $\Delta_0$.
One can then avoid solving the full
self-consistent problem by eliminating the
effective coupling constant in favor of $\Delta_0$. 
In making this useful simplification, 
one implicitly excludes small disks of radius $\sim \xi$
around each vortex in a 2D plane, where the amplitude of the order
parameter varies appreciably. This amounts to a tacit 
assumption that, in the limit $\xi\ll l$,
the effects of the detailed structure around vortex cores
on properties of the far-away nodal states in the bulk
regions are entirely negligible. 
It is this assumption that needs to be revisited, 
as we will show. Initially, however, we will follow this
assumption in order to understand the difficulties 
it generates and to introduce the notation.
%

One starts from the BdG Hamiltonian for a $d$-wave 
extreme type-II superconductor:
\begin{equation}
\label{bdg}
{\cal H}_{BdG} =
\begin{pmatrix}
\frac{(\bp-\frac{e}{c}\bA)^2}{2m}-\eps_F & \hat{\Delta}\\
\hat{\Delta^*}&\eps_F-\frac{(\bp+\frac{e}{c}\bA)^2}{2m}
\end{pmatrix}.
\end{equation}
The gap operator $\hat{\Delta}$ is given by\cite{vmft}
\begin{equation}
\hat{\Delta}= \frac1{\hbar^2 k_F^2}  \{p_x,\{p_y,\Delta(\br)\}\}  -
\frac{i}{4k_F^2}\Delta(\br)(\{\partial_x,\partial_y\}\phi),
\label{Delta}
\end{equation}
where $\Delta (\br)$ is the superconducting center-of-mass
complex gap function, $\phi$ is its phase and
curly brackets denote anticommutation operator $\{a,b \} = (ab+ba)/2$,
accompanied by the $x\leftrightarrow y$ symmetrization.
Note that the first term\cite{simonlee} in Eq. (\ref{Delta}) must
be accompanied by the second\cite{ft00,vmft}
to maintain gauge invariance of the BdG Hamiltonian -- this
amounts to replacing ordinary derivatives in the first term by
the covariant ones: ${\bf\nabla}\to {\bf\nabla}+(i/2)({\bf\nabla}\phi)$,
insuring minimal coupling of $\phi$ to the external
electromagnetic gauge potential, with charge equal to $2e$.
Although the gap operator $\hat{\Delta}$ 
in the above form has a $d_{xy}$
symmetry, the results  for $d_{x^2-y^2}$ can be 
readily obtained after a simple
rotation by $\pi/4$. 

In a uniform external magnetic field  $H$ and in the presence of
a periodic array of superconducting vortices which the field
induces in $\Delta (\br)$, both diagonal terms and 
the gap operator in (\ref{bdg}) are invariant under 
{\em  magnetic} translation group 
transformations (MTG) rather than being simply
periodic in space.  Consequently, the eigenfunctions of  
${\cal H}_{BdG}$ are not the ordinary Bloch functions but are instead
the so-called {\em magnetic} Bloch functions and can be 
classified according to representations 
of MTG \cite{brown,zak,bychkov}.
An essential aspect of the problem is the observation that a simple
Bravais lattice of superconducting vortices contains a magnetic
flux equal to $hc/2e$ per unit cell -- this is just a reflection
of the fact that a vortex is a topological defect in the phase of the gap
function of Cooper pairs which carry charge $2e$. 
In contrast, all representations of MTG in terms
of single-valued eigenfunctions must contain an integer number of
the {\em electronic} flux quanta, $hc/e$, per unit cell. It is
therefore necessary to choose the unit cell of the MTG so it
is at least twice as large as that of the vortex lattice\cite{dukan,akera}. 
Apart from this condition the shape and the size of such unit cell are 
arbitrary and different choices correspond
to different ``gauges'', i.e. different representations of MTG. The
eigenfunctions associated with these representations are different
but the spectrum of eigenvalues of ${\cal H}_{BdG}$, as measured
by its density of states, is the same for all these
choices of MTG.
  
FT observed that the above features of MTG
can be used to recast the original problem in a more convenient form.
One first partitions the original vortex lattice into two sublattices A, B
(Fig.~\ref{fig:unitcell}) and then performs a 
singular gauge transformation ${\cal H'}=U^{-1}{\cal H} U$:
\begin{equation}
U=
\begin{pmatrix}
e^{i\phi_A}&0\\
0&e^{-i\phi_B}
\end{pmatrix},
\qquad \phi(\br)=\phi_A(\br)+\phi_B(\br)~,
\end{equation}
where $\phi_{A(B)}$ are contributions to the phase of the order parameter
from  vortices of A(B) sublattice (see \cite{phiaphib} and Appendix A).
After the transformation the Hamiltonian assumes a periodic form:
\begin{equation}
{\cal H} = 
\begin{pmatrix}
\frac{(\bp+m\bv_A)^2}{2m}-\eps_F & \hat{D}\\
\hat{D}&\eps_F-\frac{(\bp-m\bv_B)^2}{2m}
\end{pmatrix}~~~~,
\label{H_FT_nonlin}
\end{equation}
where the transformed gap operator is given by
\begin{equation}
\hat{D} = e^{-i\phi_B}\hat{\Delta}
e^{i\phi_A}=
\frac{\Delta_0}{\hbar^2k_F^2}\{p_x+a_x,p_y+a_y\}
\label{fullHam}
\end{equation}
and two superfluid velocities are defined as
\begin{eqnarray}
m \bv_{A,B}(\br) &= \hbar \nabla \phi_{A,B}(\br)-e\bA/c~~,\\
\ba&=(m\bv_A-m\bv_B)/2~~.
\end{eqnarray}
Properties of the superfluid velocities
$\bv_A$ and $\bv_B$ are discussed at length further in the text and in the
Appendix. The crucial feature is that their
chirality, set by $\nabla\times\bv_{A(B)}(\br)$,
vanishes on {\em average} and thus $\bv_{A(B)}$ are 
truly periodic functions in space. Consequently, the eigenstates
of the FT transformed BdG Hamiltonian are ordinary Bloch waves that can
be classified by their crystal momentum and band index. 
In the limit of weak fields
it is reasonable to view the low energy wavefunctions
as ``perturbations'' of the nodal
states at $H=0$, and therefore the wavefunctions 
describing states  close to the nodal point $\bk^F_1=(k_F,0)$ can
be described as 
$$
\Psi(\br)=e^{i k_F x} \psi(\br),
$$
where $\psi_1(\br)$ changes weakly on distances of order $1/k_F$.
Under assumptions \mbox{$|\nabla \psi(\br)|\ll k_F |\psi(\br)|$},
\mbox{$m\bv(\br) \ll \hbar k_F$}, and
\mbox{$\hbar\nabla^2\psi \ll k_F |m\bv(\br)|$},
the Hamiltonian to the leading order is\cite{ft00}
%
\begin{equation}
\label{H_FT}
{\cal H}_{FT} = v_F (p_x+a_x)\sigma_3 +v_{\Delta} 
\sigma_1(p_y+a_y) + mv_F v_x~,
\end{equation}
where $\sigma_i$ are the standard Pauli matrices and  
$\bv=(\bv_A+\bv_B)/2$ is the gauge invariant superfluid
velocity. 
The corrections to (\ref{H_FT}) fall into two categories:
more familiar ones, of higher order in $(k_Fl)^{-1}\ll 1$ become
arbitrarily small as $H\to 0$. In addition, however,
there are corrections governed by $(k_F\xi)^{-1}\ll 1$ even as $H\to 0$,
which reflect the fact that nodes are located at {\em finite}
momentum $k_F$ and describe rapid oscillations in quasiparticle
spectra due to internodal interference\cite{vmft,foottight,tight}.

The linearized Hamiltonian ${\cal H}_{FT}$ describes a
massless Dirac particle with anisotropic dispersion in presence of
an internal gauge field $\ba(\br)$ and a scalar
potential $v_x$ given by the component of $\bv(\br)$ along the node. 
This scalar potential is nothing but the Doppler shift of nodal quasiparticle
energies in the superflow field produced by vortices.
The internal gauge field $\ba(\br)$ -- its minimal coupling to
BdG quasiparticles betraying its topological origin -- corresponds 
to an array of $\pm\frac{1}{2}$
Aharonov-Bohm fluxes located at vortex positions and generates
the required $\pm\pi$ Berry phases for fermions circling around
vortices\cite{ft00,vmft,footv}. 
This intrinsic topological frustration is
entirely absent in semiclassical approximations\cite{volovik}
but plays an essential role in the quantum mechanics of
nodal quasiparticles in presence of vortices.
Note that $\ba(\br)$ is explicitly time-reversal
invariant since the Aharonov-Bohm effect remains unaffected by
the exchange of direction of a half-flux. This is an important
feature and the necessary condition for invariance of physical results
under different FT gauge transformations. Different FT transformations
correspond to different choices of A and B sublattices (always, however,
containing the same total number of vortices so that 
$\nabla\times\bv_{A(B)}(\br)$ remains zero on average) closely resembling 
different choices of the unit cell of the MTG. Under
such transformations the Hamiltonian (\ref{H_FT}) retains its general
form but with half-fluxes in  $\ba(\br)$ switching  
their signs from positive to 
negative depending on whether they belong to  A or B sublattice, while 
$\bv(\br)$ remains the same in all FT gauges.

Hamiltonian $H_{FT}$ is periodic and has a well defined
Fourier representation. While the fact that both 
effective potentials $\ba(\br)$ and $\bv(\br)$
diverge as $1/r$ near vortices is an obvious point of concern, resulting in
matrix elements of perturbation theory that decrease only as $1/k$ 
in momentum space, one still reasonably expects that 
the path toward finding the set of eigenfunctions of $H_{FT}$
starts by simply expanding them in a plane wave basis.
\begin{figure}[tbh]
\includegraphics[width=\columnwidth]{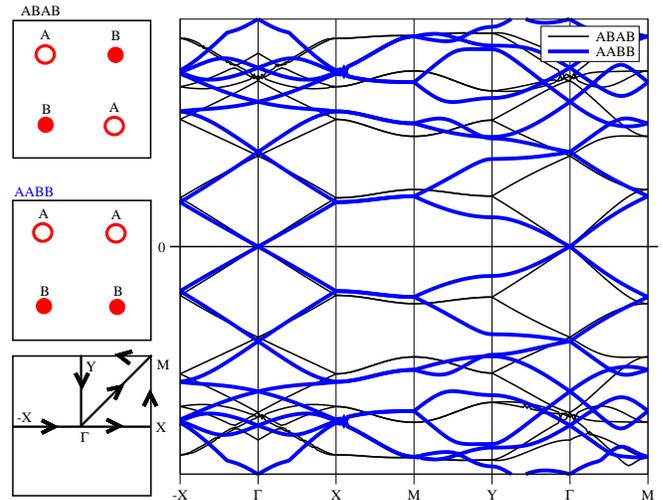}
\caption{\label{fourvortices}Comparison of energy 
spectra for different choices of A and B 
sublattices. The spectra are similar at low energies
but differ significantly as energy increases\cite{footA}.
Lower left: the first magnetic  Brillouin zone.}
\end{figure}
The numerical  solution obtained in this manner, shown in
Fig. \ref{fig:theta0.25pim0.25pi}, was found to result in
strongly dispersive bands at low energies. The low-energy
portion of the spectrum appears qualitatively similar to the zero-field 
case and, most importantly, preserves nodal point: 
$\eps(\bk)=\hbar\sqrt{v_F'^2 k_x^2+v_{\Delta}'^2k_y^2}$.
The only significant effect of finite field is 
the renormalization of the original velocities $v_F$ and
$v_{\Delta}$. The band structure calculation was 
also confirmed in\cite{marinelli}, where the spectrum and the  
eigenstates of $H_{FT}$
(\ref{H_FT}) were obtained by discretization of the Hamiltonian in real space.
The persistence of nodal points was further demonstrated\cite{marinelli,
ashvinii} to be a consequence of the general symmetry properties of $H_{FT}$.

As noted previously\cite{vmft}, the choice of 
A-B sublattice arrangement is not unique in the FT transformation, 
being merely a choice of singular
gauge. Clearly, measurable quantities such as the density of states
(DOS)  should not depend on the partitioning of vortex lattice
into various A and B sublattices.
On the other hand, the Hamiltonian $H_{FT}$ has a formal dependence on the
FT singular gauge choice, and surprisingly, the 
numerical solution exhibits small but persistent
differences for different arrangements of $A$ and $B$ sublattices (Fig.
\ref{fourvortices}). The
discrepancy remains regardless of whether the computation is done by
expanding the solution in plane wave 
basis \cite{ft00,vmft} or by discretization in real
space \cite{marinelli}. Note that these difficulties  
{\em cannot} simply be attributed to the
singular nature of the FT gauge transformation: 
the energy spectrum of the Hamiltonian
in the {\em original} representation, where the 
eigenfunctions are chosen as magnetic
translation group eigenstates,
has DOS which also exhibits a similar discrepancy 
for different choices of a magnetic unit
cell (e.g. rectangular vs. square).
As a remedy, the tight-binding formulation 
of a $d$-wave BdG Hamiltonian and FT transformation 
was introduced in\cite{vmft}, where it was verified that this so-called
``ABAB vs. AABB'' problem does not appear and the spectra are explicitly
invariant under singular gauge transformations
for arbitrary size of the tight-binding mesh. In a way, this is
the resolution of the whole ``ABAB vs. AABB''
problem: CuO$_2$ planes in high temperature
superconductors are actually well described by an effective tight-binding
Hamiltonian (of a $t-J$, Hubbard or a related variety) 
and the lattice $d$-wave
superconductor and its BdG equations is precisely what we should be
considering. The problem with the singular $1/r$ behavior never arises
since BdG quasiparticles live on sites while vortices are ``located''
in the interior of plaquettes of a tight-binding lattice.

Nevertheless, the appeal of the continuum 
formulation is undeniable.
By representing nodal quasiparticles as massless Dirac fermions
one is hoping that their interactions with vortices can be
described by a relatively simple and elegant effective continuum
theory which can then become the starting point for analytic exploration
of more complex problems involving fluctuating or disordered
vortices, thermal and quantum fluctuations of vortex-antivortex pairs, etc.
It is therefore highly desirable to understand the source of 
difficulties besetting the
continuum description of the linearized 
model and to overcome them\cite{footvi}. 
Moreover, as pointed in the Introduction, a detailed structure
of vortex cores in cuprates is unknown at
present and all possible tight-binding
approaches will necessarily suffer from being dependent on short-range
physics details, which, as one suspects, should have only a
limited effect on the nodal quasiparticle wavefunctions in the bulk
and thus on any effective low-energy theory. 
The linearized theory, in contrast, is formulated in terms of a universal
Hamiltonian (\ref{H_FT}), which is completely 
independent of short-range details, including the structure of vortex cores.

\section{Self-adjoint extensions}
\label{sae}
The origin of the difficulties with different choices of
FT singular gauge can be traced back to the assumptions made during the
linearization procedure outlined above Eq. (\ref{H_FT}). 
Although the  conditions necessary
for linearization  are
well satisfied for weak magnetic fields and in the bulk, far away
from vortex cores ($r\sim l$), they are violated
at distances $r\sim \xi$, since the neglected terms diverge as $1/r^2$
near vortices. Thus one suspects that additional regularization
might be required to take into account the effect of such terms.
Let us nevertheless assume for a moment 
that the dependence on A-B arrangement {\em is}
caused by a  numerical procedure. Since the scalar and vector potentials
$\ba(\br)$ and $\bv(\br)$ diverge as $1/r$ as one
approaches a vortex, one naturally expects that by finding the asymptotic
behavior of the eigenfunction near vortices, one can use it to his/her
advantage to cure the singularities. The asymptotics in fact were found
analytically by Melnikov \cite{melnikov}. For brevity,
in this article we will consider
mostly the isotropic case  $(v_F=v_{\Delta})$; the results are
easily generalized to the anisotropic case.

Close to a vortex two linearly independent solutions of the eigenvalue problem for
the linearized Hamiltonian  ${\cal H}_{FT}$ are described by the wavefunctions
that diverge near the vortex as $r^{-1/2}$. The angular dependence of the
divergent part of the wavefunctions $\sqrt{r}\psi(\bR_A+\br)$ near vortex A is described by
\begin{equation}
C_1e^{\frac{i\cos\varphi}{2}}
\begin{pmatrix}
e^{-i\varphi}- i\\
ie^{-i\varphi}- 1\\
\end{pmatrix}
+
C_2e^{-\frac{i\cos\varphi}{2}}
\begin{pmatrix}
e^{-i\varphi}+ i\\
ie^{-i\varphi}+ 1\\
\end{pmatrix}~,
\label{AsymptA}
\end{equation}
where $\varphi$ (not to be confused with $\phi(\br)$,
the phase of the gap function) is the polar angle around a particular vortex.
The above is only the dominant part of each eigenfunction as $r\to 0$; the
prefactors of this diverging term as well as subdominant terms are different
for different eigenfunctions. Generally, the weight of the divergent term
in an eigenfunction decreases as its energy eigenvalue increases.
%
%
The angular dependence of the singular part of the wavefunctions near
vortices of B sublattice  $\sqrt{r}\psi(\bR_B+\br)$ is given by an expression similar to
(\ref{AsymptA}):
\begin{equation}
C_1'e^{\frac{i\cos\varphi}{2}}
\begin{pmatrix}
1- ie^{i\varphi}\\
i- 1e^{i\varphi}\\
\end{pmatrix}
+
C_2'e^{-\frac{i\cos\varphi}{2}}
\begin{pmatrix}
1+ ie^{i\varphi}\\
i+ 1e^{i\varphi}\\
\end{pmatrix}.
\label{AsymptB}
\end{equation}
%
%

The asymptotics above can be checked  by retaining only the divergent parts
of the potentials  $\bv$ and $\ba$ in the FT equations with the radial
dependence proportional to $1/r$. Such ``zeroth-order'' approximation
 describes  a single vortex, since both the superfluid velocity and the vector
potential  can be represented as a sum (see Eq. (\ref{zetadef2}) of the
Appendix). Each term of the
sum represents contribution of an individual vortex, and by neglecting all
terms of the sum that are finite only the contribution of a single vortex
is taken into account. Then, one can easily verify that (\ref{AsymptA}) or
(\ref{AsymptB}), for a vortex belonging to class A or B respectively, are
exact eigenfunctions corresponding to zero energy.
Returning now to the
lattice problem, the non-singular part of the 
Hamiltonian, $\hat{V}_{reg}$, which is due to other vortices, 
can be treated as a perturbation. Such finite terms do not
modify the limiting small $r$ behavior of the zero energy eigenstates (if
there are any). Moreover, under the action of perturbation $V_{reg}$, the
nonzero energy eigenstates are mixed 
with the singular wavefunctions (\ref{AsymptA}) or
(\ref{AsymptB}), and therefore are described by the same 
limiting behavior close to any particular vortex.

The eigenfunctions of ${\cal H}_{FT}$ obtained numerically by the plane wave
expansion close to a vortex will be shown below (see
Fig. \ref{fig:FigFTthetas}) 
to have asymptotic behavior given by either (\ref{AsymptA}) or (\ref{AsymptB})
for vortices belonging to sublattice A or B, respectively.
Additionally, the increase of the wavefunctions according to power law
$r^{-1/2}$  and angular dependence given above near vortices at distances 
$\xi\ll r\ll l$ has been explicitly verified \cite{footB} 
for the solutions of  BdG equations describing the tight-binding  
lattice $d$-wave superconductor\cite{vmft,mt}.

Note that Eq. (\ref{AsymptA}) describes 
{\em two linearly independent} solutions
{\em both} of which diverge as $1/{\sqrt{r}}$ as one approaches
the vortex.
For non-singular potentials only one of the solutions
would have been square-integrable, 
and the second solution would have to be discarded.
Thus, before one proceeds, one must decide
whether such divergent solutions are permitted in the spectrum.

\subsection{Dirac particles in the field of Aharonov-Bohm flux}
This issue is not entirely new, and similar
problems were encountered in the context of
cosmic strings \cite{sousajackiw}. There, a Dirac
particle is considered in the field of a single Dirac string carrying a
fractional magnetic flux. We will restrict 
our discussion to the relevant case of a single half-integer flux:
\begin{equation}
\begin{pmatrix}
p_x+a_x & p_y+a_y\\
p_y+a_y & -p_x-a_x
\end{pmatrix}
\begin{pmatrix}
u\\
v
\end{pmatrix}
=
E
\begin{pmatrix}
u\\
v
\end{pmatrix}~,
\label{fluxprob}
\end{equation}
where $\ba=\frac{(-y,x)}{2r^2}$ is the vector potential corresponding to
the Aharonov-Bohm half-flux. The Hamiltonian can be obtained
from (\ref{H_FT}) where the last term describing scalar potential is set to
zero. 
After rotation  $\chi_1=u-iv$, $\chi_2=u+iv$ we obtain Hamiltonian in the
basis used in \onlinecite{sousajackiw}:
\begin{equation}
H_{GJ}=
\begin{pmatrix}
0 & p_x+a_x -i(p_y+a_y)\\
p_x+a_x+i(p_y+a_y) &0
\end{pmatrix}~.
\label{chiral}
\end{equation}
The eigenstates of $H_{GJ}$ can be 
expressed through Bessel functions as 
\begin{equation}
\begin{pmatrix}
\chi_1\\
\chi_2
\end{pmatrix}
=e^{in\varphi}\left[
C_1\begin{pmatrix}
J_{n+1/2}\\
ie^{i\varphi}J_{n+3/2}
\end{pmatrix}+
C_2
\begin{pmatrix}
J_{-n-1/2}\\
-ie^{i\varphi}J_{-n-3/2}
\end{pmatrix}\right]~.
\label{gensol}
\end{equation}
Gerbert and Jackiw noticed that for all angular channels except
$n=-1$ the square integrability requirement specifies
which of the two independent solutions in (\ref{gensol}) should be present.
For $n=-1$, however, there is an ambiguity 
as both solutions are square integrable, but divergent as
$1/{\sqrt{r}}$ at the origin. They found  
that keeping both wavefunctions in the spectrum
results in an overcomplete basis, with such pathologies as imaginary
eigenvalues of the Hamiltonian. 
On the other hand, the requirement of continuity (non-divergence) of the
wavefunctions at the origin is too strong and causes
the loss of completeness in the angular momentum channel.

The complications are due to the singular nature of the vector potential
$\ba(\br)$ at the origin. The usual spectral 
theorems, normally derived for {\em
bounded} symmetric (Hermitian) operators, do not hold automatically for
singular potentials and a more careful 
analysis is needed. The domain of finite
wavefunctions should be enlarged to include a
specific linear combination of $1/{\sqrt{r}}$-divergent solutions. Using the 
language of functional analysis, self-adjoint extensions of the Hamiltonian
should be constructed~\cite{reed}. Gerbert and Jackiw found that the
mathematically allowed linear combination of divergent wavefunctions is
not unique but forms a one-parameter family
\begin{multline}
\chi\propto
\sin\theta
\begin{pmatrix}
J_{-1/2}(Er)\\
ie^{i\varphi}J_{1/2}(Er)
\end{pmatrix}+
\cos\theta
\begin{pmatrix}
J_{1/2}(Er)\\
-ie^{i\varphi}J_{-1/2}(Er)
\end{pmatrix}\\
\sim \frac1{\sqrt{r}}
\begin{pmatrix}
\sin\theta\\
-i e^{i\varphi}\cos\theta
\end{pmatrix}~~,
\label{fluxsae}
\end{multline}
where $\theta\in[0,\pi)$ is a real parameter that labels distinct 
self-adjoint extensions.
The parameter $\theta$ expressing the boundary condition cannot be found
from the model that treats the string as a ``black box''; it depends on the
short-scale structure of the string. 

A string described by a divergent vector potential at $r=0$ is of course
only an idealization. To find appropriate $\theta$ one has to consider the
physical regularization of the problem. The simplest
case of magnetic field concentrated in a thin 
cylindrical shell of small, but finite radius
$\eps$ when $\eps\to 0$ was considered 
in \onlinecite{alford}. By matching solutions
inside and outside the core, the authors
found that for radially extended symmetrical distribution of magnetic field
inside the core one obtains $\theta=0$, implying that
the lower component of spinor $\chi$
stays regular at the origin. The procedure can be repeated for physical,
extended fluxes carrying arbitrary half-integer flux $\Phi$, and 
it  was found that for $\Phi=1/2,3/2,5/2,7/2{\ldots}$ 
the self-adjoint extension
is the same  $\theta=0$ and the solutions for positive half-integer fluxes
are related to each other by singular gauge 
transformations $\chi(\br)\to\exp(iN\varphi) \chi(\br)$. Quite 
surprisingly, the negative extended half-integer fluxes
$\Phi=-1/2,-3/2,{\ldots}$ belong to a different class $\theta=\pi/2$ which
corresponds to wavefunctions with regular upper components of $\chi$. Thus,
the symmetry $\Phi\to \Phi+1$ is broken around $\Phi=0$. 
Due to the unbounded increase
of the wavefunctions, a relativistic fermion
can ``probe'' the short-scale structure of the core
and the direction of the flux is ``exported'' to the exterior through boundary
condition\cite{alford}. Although physical, extended fluxes do have
different limiting solutions for opposite orientations of magnetic field 
as the size of the core is taken to zero, one can still perform singular
gauge transformations, of course, provided that the boundary condition
$\theta$ is chosen correctly to reflect the 
direction of the real, physical magnetic field.

A necessity of considering self-adjoint extensions is not restricted to
a single-flux problem. In the case of two fluxes the problem can also
be solved essentially exactly; the solution is presented in the
Appendix B.

\subsection{Dirac particle in the field of 
periodic array of singular scatterers}

The above analysis of a single flux 
problem\cite{sousajackiw} is facilitated due to the knowledge of exact
eigenfunctions allowing for application of the von Neumann theorem
on self-adjoint extensions\cite{reed}.
To consider an array of vortices, 
we will rederive the result (\ref{fluxsae})
by  following a different procedure, which will 
allow us to find the self-adjoint
extensions corresponding to
the vortex lattice problem, for which exact eigenfunctions are
unknown. We consider a general Hamiltonian
\begin{equation}
H=
\begin{pmatrix}
p_x+V_{11} & p_y+V_{12}\\
p_y+V_{21} & -p_x+V_{22}
\end{pmatrix}~~,
\end{equation}
where $V_{ij}$ are arbitrary periodic functions which can diverge at most
as $1/r$ at a certain finite
set of points in unit cell, such as positions of fluxes
or vortices.


In order for  Hamiltonian (\ref{H_FT}) to be 
a symmetric  operator $(\psi^* H \chi) = (\chi^* H\psi)^*$, the condition
\begin{equation}
\label{SymmDef}
\int\left[ \partial_x (\chi_1\psi_1^* - \chi_2 \psi_2^*)
+\partial_y(\chi_1\psi_2^*+\chi_2\psi_1^*)\right] d^2 r=0
\end{equation}
must be fulfilled. This is satisfied 
automatically if $\chi(\br)$ and $\psi(\br)$
have different crystal momenta, and therefore we concentrate on the
subspace of the wavefunctions with the same crystal momentum $\bk$, but
different band indices.
Symmetric operator is self-adjoint if the domain $D(H)$ coincides with that
of its adjoint $D(H^*)$. In other words, we have to find such a boundary
condition on $\psi(\br)$ so that the adjoint wavefunctions  $\chi(\br)$ 
satisfy precisely the same condition.
\psfrag{figFT_A_1}{$|u(r)|\sqrt{r}$}
\psfrag{figFT_A_2}{$|v(r)|\sqrt{r}$}
\psfrag{figFT_A_3}{$|u(r)|/|v(r)|$}
\psfrag{figFT_A_4}{$r$}
\psfrag{figFT_A_5}{ $\lim_{r\rightarrow
0}\left|\frac{u(\br)}{v(\br)}\right|$ }
\psfrag{figFT_A_6}{$\theta$}
\begin{figure}
\includegraphics[width=3.4in]{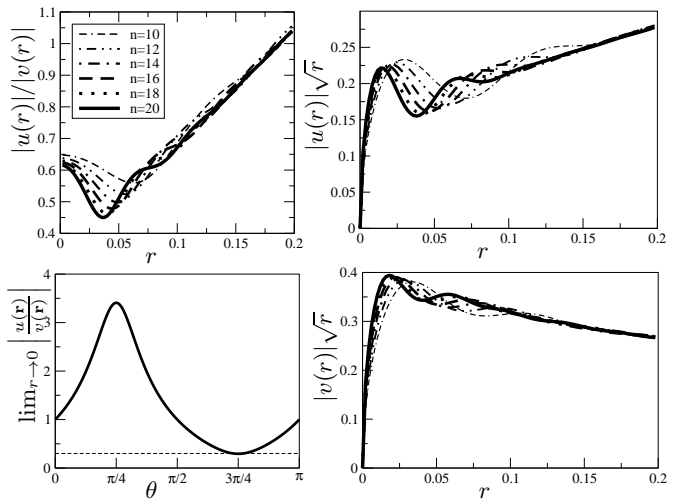}
\caption{\label{fig:FigFTthetas}
Typical eigenfunctions of the linearized equation found by expanding the
wavefunctions in plane wave basis without explicitly imposing boundary
conditions at vortex locations. $N=(2n+1)^2$ plane
waves are included in numerical solution.
Right panels:  $\sqrt{r}|u(r)|$ and
$\sqrt{r}|v(r)|$ along fixed direction $\varphi=0$  
as functions of distance from
the vortex. Each plot approaches a straight line as the accuracy is
increased. The convergence is non-uniform exhibiting Gibbs overshoot as
expected, because of the singularity of the wavefunction at the origin. 
Upper left panel: ratio $\left|u(r)/v(r)\right|$ used to extract $\theta$.
The intercept of the limiting linear dependence ($0.3+0.67 r$) is used to find $\theta$ as
shown in the lower left panel.
}
\end{figure}
For the subspace of functions characterized by the same crystal momentum
$\bk$, the integrand of (\ref{SymmDef}) is periodic 
in space, and therefore the integral 
over entire space can be replaced by an integral over one
unit cell. It may appear that the integral automatically vanishes;
by Stokes theorem it can be transformed 
into a contour integral along the edges of the unit
cell, which in turn
vanishes due to the periodicity of the integrand. Indeed, this
conclusion is valid for {\em regular} wavefunctions.
It must be remembered, however, that the necessary condition for Stokes
theorem to hold is the continuity
of the partial derivatives in the integrand in (\ref{SymmDef}) inside the
contour, and in our case, with $\psi\propto 1/\sqrt{r}$ near a 
vortex, the theorem does not automatically apply. 
To proceed with the analysis, we define
\begin{equation}
\left\{
\begin{aligned}
\cA_x&=\chi_1\psi_2^*+\chi_2\psi_1^*\\
\cA_y&=\chi_2\psi_2^*-\chi_1\psi_1^*
\end{aligned}
\right.
\end{equation}
and separate the contribution of singularities near the vortices by
drawing  discs $\Gamma_i$ of radii $\eps_i$ surrounding each
vortex (see Fig.~\ref{fig:unitcell}(b) for 
the case of two  vortices per unit cell).

After Stokes theorem is applied to the exterior of the discs
within the unit cell, where the wavefunctions are finite, 
the contribution from the edges cancels by
periodicity, and we find that the sum of the contour integrals along the
boundaries of the discs $\Gamma_i$
and the area integrals over the interior of $\Gamma_i$ equals
\begin{equation}
\sum_{i} \int_0^{2\pi}  (\sin \varphi A_y(\eps_i,\varphi) 
- \cos\varphi A_x(\eps_i,\varphi))d\varphi = 0~~,
\label{SAEreq}
\end{equation}
where $i$ labels the vortices in unit cell and $\varphi$ is the polar angle
around the $i$th vortex.
%
%
Obviously, if one demands that $\psi(\br)$ are regular at the origin,
then the adjoint domain of $\chi$ includes all solutions that behave as
$1/\sqrt{r}$, leading to $D(H)\neq D(H^*)$.
Now note that in the limit  $\eps_i\rightarrow 0$ 
only the singular part of the
wavefunctions (\ref{AsymptA}) and (\ref{AsymptB}) contributes.
Since the limits $\eps_i\rightarrow 0$ can be taken independently,
it is straightforward to find from (\ref{SAEreq})
that the self-adjointness requirement
$D(H)=D(H^*)$ fixes the relative phase between two asymptotic
solutions\cite{theothersae}.
When applied to the array of fluxes we find that this requirement
leads to the same boundary conditions as in
\onlinecite{sousajackiw}. In terms of $(u,v)$ these boundary conditions in the
vicinity of fluxes are
\begin{align}
\sqrt{r}\begin{pmatrix}
u\\
v
\end{pmatrix}&\rightarrow
\cos\theta
e^{-i\phi}
\begin{pmatrix}
1\\
i
\end{pmatrix}
+
\sin\theta
\begin{pmatrix}
i\\
1
\end{pmatrix}
&&\text{for $+\frac{\Phi_0}2$ fluxes}\\
\sqrt{r}\begin{pmatrix}
u\\
v
\end{pmatrix}&\rightarrow
\cos\theta
\begin{pmatrix}
1\\
i
\end{pmatrix}
+
\sin\theta
e^{i\phi}
\begin{pmatrix}
i\\
1
\end{pmatrix}
&&\text{for $-\frac{\Phi_0}2$ fluxes}~.
\end{align}
These boundary conditions are actually equivalent to those of
Ref. \onlinecite{sousajackiw}.

For the present problem of $d$-wave quasiparticles 
in a vortex lattice the above
requirement of self-adjointness $D(H)=D(H^*)$ translates into the
following boundary conditions near A vortices:
\begin{multline}
\sqrt{r}\psi(\bR_A+\br)
\rightarrow\\
\cos\theta e^{\frac{i\cos\varphi}{2}}
\begin{pmatrix}
e^{-i\varphi}- i\\
ie^{-i\varphi}-1\\
\end{pmatrix}
+
\sin\theta e^{-\frac{i\cos\varphi}{2}}
\begin{pmatrix}
e^{-i\varphi}+i\\
ie^{-i\varphi}+1\\
\end{pmatrix}.
\label{bcA}
\end{multline}
The boundary conditions at B vortices differ by the overall phase
factor $\exp(i\varphi)$: 
\begin{multline}
\sqrt{r}\psi(\bR_B+\br)
\rightarrow\\
\cos\theta e^{\frac{i\cos\varphi}{2}}
\begin{pmatrix}
1-ie^{i\varphi}\\
i-e^{i\varphi}\\
\end{pmatrix}
+
\sin\theta e^{-\frac{i\cos\varphi}{2}}
\begin{pmatrix}
1+ie^{i\varphi}\\
i+e^{i\varphi}\\
\end{pmatrix}.\\
\label{bcB}
\end{multline}

Thus, if the interior of a vortex is 
treated as a black box, in order to fully specify the problem, 
the linearized Hamiltonian $H_{FT}$ must be supplemented by 
boundary conditions at the locations of vortices\cite{vmft}. Every boundary
condition depends on a single parameter $\theta$, which is easily shown
to be independent of an A-B assignment.
Indeed, the transformation from one 
choice of A and B sublattices to another is
given by a unitary matrix
$$
U_{AA'}=
\begin{pmatrix}
e^{i\phi_A-i\phi_{A'}}&0\\
0&e^{i\phi_B'-i\phi_{B}}
\end{pmatrix}=
e^{i\phi_A-i\phi_{A'}}\cdot 1~~,
$$
since $\phi_A+\phi_B=\phi_{A'}+\phi_{B'}=\phi(\br)$. $U_{AA'}$ is
proportional to unit matrix, and the asymptotic behavior of the wavefunctions,
specified by $\theta$, is clearly invariant under all such
transformations. Note that parameter $\theta$ for a given vortex
has been intentionally
defined in this particular
way for convenience. It is independent on whether this individual vortex
belongs to an A  or a B sublattice, and is therefore a scalar
under FT singular gauge transformations. Naturally, one must recall here
that the actual form of the boundary
condition {\em does} change according to Eqs. (\ref{bcA},\ref{bcB})
depending on whether a particular vortex belongs
to an A or a B sublattice. Furthermore, note that there is 
no {\em a priori} requirement that
parameters $\theta$ are equal for different vortices within unit cell.
Just as in the case of a single Dirac string, parameters $\theta$ cannot be
determined from the linearized Hamiltonian. Rather, the boundary conditions
$\theta$ are determined by the short-ranged physics of fully self-consistent
BdG equations and include all effects
of higher orders in $(k_F l)^{-1}$ expansion.

Equipped with this new understanding we 
now revisit the ``ABAB vs. AABB'' problem
and consider the 
numerical computations performed earlier
\cite{ft00, vmft}. There boundary conditions were not explicitly enforced
and the numerical procedure itself ``spontaneously'' selected 
a particular set of boundary conditions. 
In calculations done in \onlinecite{ft00,vmft}, 
the wavefunctions are represented
as linear combinations of plane waves, and, 
after a substitution into the linearized FT Hamiltonian, the
series is truncated at some number 
of terms $N=(2n+1)^2$, which is gradually increased until
convergence is achieved:
\begin{equation}
\psi_{\bk,n}(\br) = e^{i\bk \br}\sum_{\bG\in \Sigma_N} e^{i\bG \cdot \br} 
\begin{pmatrix}
u_{\bG}\\
v_{\bG}
\end{pmatrix}~.
\end{equation}
Rather than analyzing the spectra we first concentrate on the
wavefunctions. The solutions $\psi(\br)=(u(\br),v(\br))$ are shown in
Fig.~\ref{fig:FigFTthetas}.
The wavefunctions indeed approach  $1/\sqrt{r}$ 
dependence on the distance from 
the vortex as the number  of reciprocal lattice vectors included in the
solution is increased. The values of $\theta$ for $\psi_n(\bk)$ can be
extracted from the wavefunctions by various methods. As an example,
the procedure involving 
ratios $\lim_{\rho\rightarrow 0} |u(\rho,\varphi=0)|/|v(\rho,\varphi=0)|$ 
is illustrated in Fig.~\ref{fig:FigFTthetas}. Since there are two vortices
per unit cell\cite{ft00} 
 we label them ``A'' and ``B'' to indicate their belonging to
two different sublattices -- to avoid any confusion,
it is important to stress here that
the labels ``A'' and ``B'' refer to 
{\em two physically distinct} vortices rather
than to an FT gauge label attached to an individual vortex (the quotation
marks are used to highlight this difference).
The value $\theta$ for
vortex ``A'' turns out to be close to $\theta=(0.75\pm 0.03)\pi$, while vortex
``B''  has asymptotic form with  $\theta=(0.25\pm0.03)\pi$. 
The conclusion remains valid not only for the vicinity of nodal points,
but also for all bands at arbitrary $\bk$ for which $\theta$'s could be
extracted. 
As discussed in a separate paper,\cite{tight}
these values of $\theta$'s
are among several consistent with the 
general symmetry requirements that must be
obeyed by the solutions of the full BdG equations. They
thus faithfully represent a
possible solution for the low energy spectrum of BdG equations. 
The meaningful determination of $\theta$ for high energy bands 
was beyond the precision of our calculations 
since the contribution of the singular
part of the wavefunctions decreases for higher bands, and one has to
consider distances that are extremely
close to the vortex to extract the asymptotic
$1/\sqrt{r}$ behavior. This, in turn requires diagonalization of matrices
with unrealistically large $N$.

We then performed a similar analysis of 
several other arrangements of A and B sublattices.
We  found that the eigenstates of ``AABB'' and ``ABAB'' lattices for
four vortices per unit cell and similar vortex lattices
rotated by $\pi/4$ with respect to the nodal 
directions also exhibit the same pattern:
the plane wave expansion procedure drives ``A'' vortices to
$\theta\approx 0.75\pi$ and ``B''  vortices to
$\theta\approx 0.25\pi$. 
The discrepancies encountered earlier that gave rise
to the ``ABAB vs. AABB'' problem\cite{vmft} 
are now easy to understand, since
the problems solved without externally 
imposing boundary conditions at vortex sites
corresponded to physically {\em different} situations once boundary
conditions were ``spontaneously''  generated by the chosen procedure,
i.e. the two problems solved were {\em not connected} by an FT
singular gauge transformation. 
In particular, the ``ABAB'' sublattice partition corresponds to
the boundary conditions at vortex locations with $\theta=\pm \pi/4$ assigned 
in a checkerboard arrangement.  In contrast, if vortices ``A'' and ``B''
are chosen to form series of parallel lines, as in the 
``AABB'' partitioning, the 
numerical procedure spontaneously selects boundary conditions
arranged as interchanging lines of $\theta=\pi/4$ and $\theta=-\pi/4$.
Obviously, these are two genuinely different physical situations in
light of the condition that $\theta$ for a particular vortex must
remain unchanged under FT gauge transformations.

It is clear that the boundary conditions at vortex locations must be fixed
separately and changed appropriately under FT transformation following
Eqs. (\ref{bcA},\ref{bcB}), as anticipated in \onlinecite{vmft}. A specific $\theta$ 
assignment can be implemented by a procedure which represents a 
modification of a plane wave expansion 
similar to the orthogonalized plane waves
method. This procedure is introduced and discussed 
in detail in section V and yields results in 
mutual agreement for different FT singular gauge choices.
Alternatively, the boundary conditions can be emulated by adding
a mass term to the linearized Hamiltonian that grows rapidly
as one approaches a vortex core. The latter approach is described
in section VI.

\begin{table}[tbh]
\noindent\begin{tabular}{|l|l|}
\hline
Group element $h$&
Transformation of $\bv$ and $\ba$\\
\hline
\hline
\noindent\begin{tabular}{l}
Mirror symmetry $m_x$:\\
$m_x\br=(l/2-x,y)$
\end{tabular}
&
$ 
\begin{array}{ll}
v_x(g\br) = v_x(\br)&v_y(g\br) =-v_y(\br)\\
a_x(g\br) =a_x(\br)&a_y(g\br) = -a_y(\br)
\end{array}
$
\\
\hline
\begin{tabular}{l}
Mirror symmetry $m_y$:\\
$m_y\br=(x,l/2-y)$
\end{tabular}
&
$ 
\begin{array}{ll}
v_x(g\br) =- v_x(\br)&v_y(g\br) =v_y(\br)\\
a_x(g\br) =-a_x(\br)&a_y(g\br) = a_y(\br)
\end{array}
$\\
\hline
\noindent\begin{tabular}{l}
Inversion $I=m_xm_y$:\\
$I\br=2\bR_A-\br$
\end{tabular}
&
$
\begin{array}{l}
\bv(g\br) = -\bv(\br)\\
\ba(g\br) = -\ba(\br)
\end{array}
$
\\
\hline
\noindent\begin{tabular}{l}
Inversion $A\to B$\\
$P\br= -\br$
\end{tabular}
&
$
\begin{array}{l}
\bv(g\br) = -\bv(\br)\\
\ba(g\br) = \ba(\br)
\end{array}
$\\
\hline
$P m_x\br=(x+l/2,-y)$&
$ 
\begin{array}{ll}
v_x(g\br) = -v_x(\br)&v_y(g\br) =v_y(\br)\\
a_x(g\br) =a_x(\br)&a_y(g\br) = -a_y(\br)
\end{array}
$
\\
\hline
$P m_y\br=(-x,y+l/2)$&
$ 
\begin{array}{ll}
v_x(g\br) = v_x(\br)&v_y(g\br) = -v_y(\br)\\
a_x(g\br) = -a_x(\br)&a_y(g\br) = a_y(\br)
\end{array}
$
\\
\hline
\noindent\begin{tabular}{l}
Translation $s=PI$:\\
$s\br=\br+2\bR_A$
\end{tabular}
&
$\begin{array}{l}
\bv(g\br) = \bv(\br)\\
\ba(g\br) = -\ba(\br)
\end{array}
$\\
\hline
\end{tabular}
\caption{Symmetry properties of superfluid velocity $\bv$ and 
internal gauge field
$\ba$.}
\label{symvAB}
\end{table}

\section{Symmetries of the single-node Hamiltonian}
\label{sym}
In the previous section we showed that the linearized
FT Hamiltonian $H_{FT}$ must be supplemented by boundary conditions
at vortex locations. These boundary conditions are fixed by
a dimensionless parameter $\theta\in [0,\pi)$. The specific 
value of $\theta$ that should be used is determined by the physics
unfolding at the lengthscales shorter than those included in the linearized
description. Thus, we are at an impasse; we apparently must solve
for the physics beyond linearization to fully specify the linearized
problem itself. We take the initial step toward such a solution
in a separate paper,\cite{tight} where we study a lattice
$d$-wave superconductor. In the present paper, we continue our analysis
of different
$\theta$'s by focusing on
the symmetry properties of $H_{FT}$ in combination with those of
the original, non-linearized BdG problem (\ref{bdg},\ref{H_FT_nonlin}).  

Certain aspects of the symmetry properties of the linearized Hamiltonian
$H_{FT}$ were already considered previously\cite{marinelli, ashvinii}. 
For example, it has been
shown that the Dirac node of the zero-field problem is not destroyed by 
the inversion-symmetric vortex lattice; the only situation considered,
however, is the one with no boundary conditions imposed
on the vortices.  Here, we extend and generalize these early results by
analyzing the complete group of symmetry 
transformations of $H_{FT}$ and examine
the consequences of imposing the boundary conditions (\ref{bcA},\ref{bcB}).
For convenience, only the simplest choice of a unit cell
(Fig.~\ref{fig:unitcell}) containing two
vortices per unit cell will be studied.
Using the  Fourier representation  
(\ref{superv1}) and (\ref{superv2}), the superfluid velocity
$\bv(\br)$ and the internal gauge field $\ba (\br)$ can be shown to transform
under geometric point transformations $h$ according  to Table \ref{symvAB}.
Note that under inversion $I$ around a vortex both $\bv(\br)$ and $\ba(\br)$
change sign, while the inversion operation  $P$ around the midpoint
between vortex A and vortex B which interchanges vortices A and B,
leaves the internal gauge field $\ba(\br)$ invariant: $\ba(-\br)=\ba(\br)$.

The operations $h$ do not form a group by themselves, since their
product can generate a pure translation by a lattice vector. Rather, the
group is formed by elements $T_{\bR}h$ 
where $T_{\bR}$ is an arbitrary translation by
a lattice vector $\bR$. One can easily check that the group can be generated
by forming products of primitive translations, 
reflections $m_x$, $m_y$ and an inversion
$P$. The symmetry of the potentials $\bv(\br)$ and $\ba(\br)$ is higher than
the group just described,  including
also operations generated by rotations by $\pi/4$ around a vortex;
these additional operations, however, 
do not correspond to any symmetry of the linearized Hamiltonian
due to the last term in (\ref{H_FT}) reflecting the fact that the boost
in the direction of one of the nodes breaks the four-fold symmetry of the
original BdG Hamiltonian even in 
the isotropic case $v_F=v_\Delta$. 
Only after the contributions of all four 
nodes are combined, will this symmetry be fully restored.

It should be emphasized that the symmetry properties discussed in this
section refer to the linearized Hamiltonian $H_{FT}$ and its eigenfunctions
$\psi(\br)$. $\psi(\br)$ are related to the 
wavefunctions $\Psi(\br)$ in the original BdG basis
according to 
$$
\Psi(\br) = e^{ik_F x}
\begin{pmatrix}
e^{i\phi_A(\br)} &0\\
0&e^{-i\phi_B(\br)}
\end{pmatrix}
\psi(\br).
$$
All transformations introduced above are understood as {\em intra}-nodal. 
For example,  inversion  $P$ transforms  $\psi(\br)$ to $\psi(-\br)$, and
the transformed wavefunction in the original basis is given by
\begin{equation}
P \Psi(\br) = e^{ik_F x}
\begin{pmatrix}
e^{i\phi_A(\br)} &0\\
0&e^{-i\phi_B(\br)}
\end{pmatrix}
\psi(-\br)~~.
\label{blah}
\end{equation}
Properties of the original non-linearized 
Hamiltonian (\ref{H_FT_nonlin}) under various
symmetry operations and their relation to the symmetries of ${\cal H}_{FT}$
will be discussed separately in section \ref{relation}.

Before analyzing the problem with imposed boundary conditions, we first
consider formally the 
linearized Hamiltonian ${\cal H}_{FT}$ with no boundary
conditions enforced at vortex locations. 
Although such ``unrestricted'' problem is in principle not guaranteed
to represent any particular 
physically meaningful linearization of the full BdG Hamiltonian (\ref{bdg}), 
this discussion will serve as the basis for subsequent analysis.
After the boundary conditions are enforced, 
the allowed symmetry operations will
form a subset of the full transformation group of the ``unrestricted''
Hamiltonian, which depends on the choice of the self-adjoint extension.

Using the  properties of superfluid 
velocities from Table \ref{symvAB} it is straightforward to verify
that  if $\Psi_{\bk}$ is an eigenfunction of 
Hamiltonian ${\cal H}_{FT}$ with crystal
momentum $\bk$ and eigenvalue $E$ then 
$\Psi'_{\bk'}=D(g)\Psi_{\bk}$ is an eigenfunction of ${\cal H}_{FT}$ with
crystal momentum and eigenvalue $\pm E$ according to the
Table \ref{sympsi} in the Appendix. In identifying $\bk'$ after the 
transformations involving phase factor
$$
f(\br)\equiv\exp(-i\delta\phi)\equiv\exp(i\phi_B-i\phi_A)
$$
it is important to keep in mind that $f(\br)$
is not periodic but anti-periodic -- it changes 
sign under primitive translations along $x$ and
$y$ directions (see Appendix): 
\begin{align}
f(x+l,y)&=-f(x,y)\\
f(x+l,y)&=-f(x,y)
\end{align}
Transformations $D(g)$ form  
ray representation\cite{footRay} of the group, and the standard 
methods can be applied to analyze the degeneracy of the eigenstates.
In addition to geometric transformations and 
pure translations $D(T_{\bR})$, there
is an additional anti-linear complex-conjugation transformation:
$$
D(C) \Psi = i e^{-i\dphi}\sigma_2 \overline{\Psi}
$$
$D(C)$ commutes with all other operators up to a phase factor, and 
combined with  the operations listed in the 
left portion of Table \ref{sympsi} from Appendix, $C$ generates 8
new eigenfunctions with energy $\pm E$.
Since Bloch basis has already been
 chosen to diagonalize ordinary translation
operators we have the total of
$16$ non-trivial transformations, which generate $8$
eigenfunctions with energy $E$ and $8$ eigenfunctions with energy $-E$.

\begin{figure}
\includegraphics[width=0.8\columnwidth]{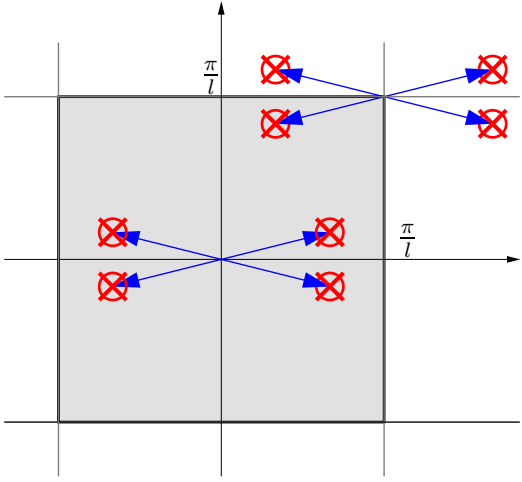}
\caption{\label{fig:unrestricted}First Brillouin zone is shown with the 
star of wavefunctions generated by applying symmetry operations
to $\Psi_{\bk}(\br)$. For convenience both, positive energy states (open
circles) and negative energy eigenfunctions (crosses) are shown.
}
\end{figure}

\begin{figure*}
\includegraphics[width=\textwidth,clip]{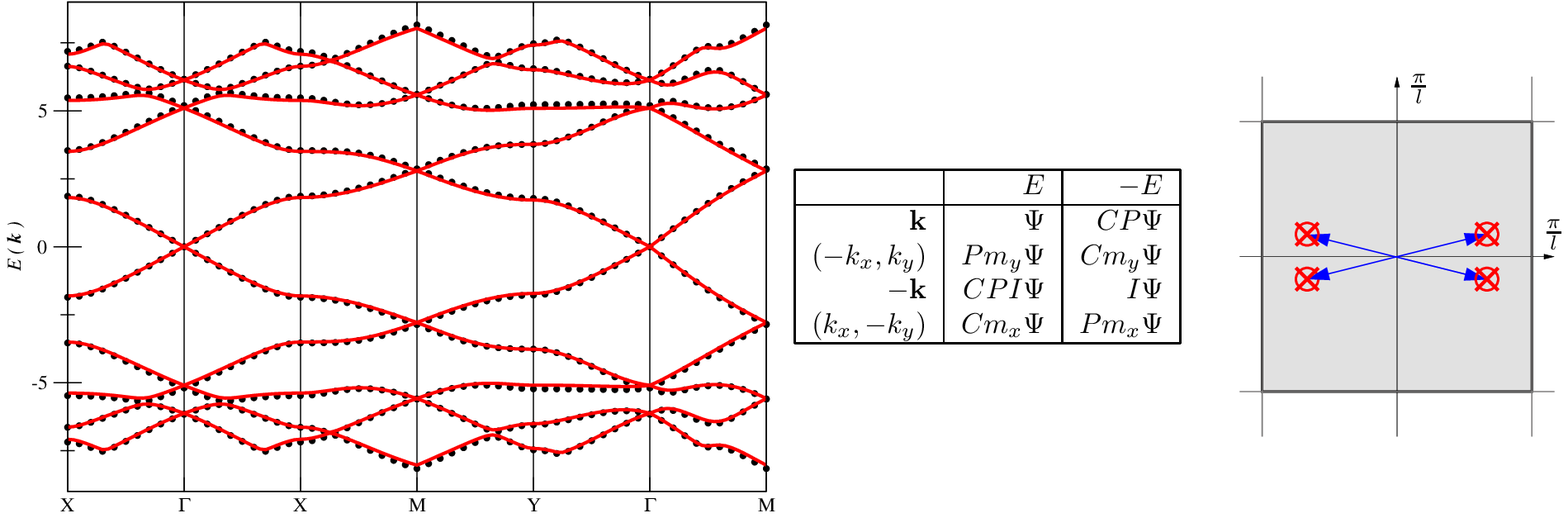}
\caption{\label{fig:theta0.25pim0.25pi}Left panel: the 
energy bands of linearized Hamiltonian for two
vortices per unit cell calculated by plane wave expansion\cite{ft00} (solid
lines) and by OPW method with boundary conditions 
$\theta_A=-\pi/4$ and $\theta_B=+\pi/4$ (dots).  
Center panel: symmetry operations relating
eigenenergies at different $\bk$. Right panel: the first Brillouin zone
with equivalent $\bk$ vectors shown by  circles. For every state with energy
$E$ at momentum $\bk$ there is a state with the same momentum and energy
$-E$.}
\end{figure*}

If $\bk$ is an arbitrary point of the Brillouin zone that does not lie on
one of the symmetry lines, all $8$ states with energy $E$
correspond to distinct crystal momenta $\bk'$  forming
a ``star'' shown in Fig.~\ref{fig:unrestricted}, and are therefore
linearly independent.
Consequently, for general $\bk$ the symmetry of the Hamiltonian is not
sufficient to result in degeneracy of the bands at the same $\bk$. For every
state $(E,\bk)$ there is a state with the opposite energy $(-E,\bk)$
obtained by applying transformation $CP$ to the original wavefunction and
seven other pairs of states with energies $\pm E$ with crystal momenta $\bk'$
belonging to the ``star''. For certain particularly
symmetric values of  $\bk$ two or more
points of the ``star'' coincide, and therefore the spectrum may become
degenerate, as will be discussed shortly in detail after the boundary
conditions are chosen.
Note, however, that the Hamiltonian apparently possesses higher symmetry
than what is reflected in the band structure spectrum shown in 
Fig.~\ref{fig:theta0.25pim0.25pi}, which was obtained numerically by the 
plane wave expansion (without
imposing boundary conditions explicitly 
near vortices)\cite{ft00, marinelli,vmft}.
In particular, the symmetry analysis indicates that the node
at $\bk=0$ implies also the identical node centered around the corner of the
Brillouin zone $\bpi$, while no such additional nodes are seen in numerical
computation. In view of our previous discussion of the role of boundary
conditions, the discrepancy is easy to understand: we have shown that, even
if the boundary conditions at the vortices are not explicitly enforced from
the outset, the numerical procedure selects a certain self-adjoint
extension, which has the effect of reducing the symmetry of the problem. 

To study the effect of boundary conditions on the symmetry of the
spectrum, we note that, if one starts with Hamiltonian ${\cal H}_{FT}$ with
boundary conditions on the wavefunctions $\psi(\br)$ given by parameters $\theta_{1(2)}$ near the two
vortices, then after general transformation wavefunctions $D(g)\psi(\br)$
have asymptotics  described by a set of 
different boundary conditions $\theta'_{1(2)}$.
The new parameters $\theta'_{1(2)}$ are shown in the rightmost column of
Table \ref{sympsi} from Appendix. Note
that the spectrum is symmetric under $(\bk,E)\rightarrow (\bk,-E)$ and
only if $\theta_1=-\theta_2$. In the case when
$\theta_1=\theta_1$ that leads to only two possible choices:
$\theta_1=\theta_2=0$ and $\theta_1=\theta_2=\pi/2$. For these extensions,
however, the inversion property $E(\bk)=E(-\bk)$ is lost and we
find  that no self-adjoint extension exists with the spectrum resembling
that of the zero-field Dirac particles, if all  
vortices are assigned the same value of $\theta$.


{\em Plane wave extension.} The numerical analysis of
section \ref{sae} showed that the straightforward plane wave expansion
``spontaneously'' generates solutions with the boundary 
conditions $\theta_{1(2)}\approx \pm \pi/4$.
We now first demonstrate that the spectrum of the self-adjoint extension
with  $\theta_{1(2)}=\pm \pi/4$ indeed has the symmetry properties
consistent with  the numerical solution
and then clarify why the plane wave expansion procedure favors these
particular boundary conditions.

\psfrag{pioverl}{\large $\frac{\pi}{l}$}

The subset of symmetry operations that 
leaves $\theta_{1(2)}=\pi/4$ invariant
and  its star of the $\bk$-vectors is 
shown in Fig.~\ref{fig:theta0.25pim0.25pi}.
The special points in BZ where the spectrum can 
potentially become degenerate are
$\Gamma$, $X$, $Y$, and $M$, where all four vectors of the star
coincide \cite{foot_symlines}. Whether
this indeed leads to degenerate eigenfunctions
depends on $\bk$ and should be analyzed in each case
separately. Using  Table \ref{sympsi} from Appendix, 
the representations matrices
$D(PM_y)=g_1$, $D(CPI)=g_2$, and $D(Cm_x)=g_3$  satisfy
\begin{equation}
\begin{array}{lll}
g_1^2 = e^{ik_yl} & g_1 g_2 =i e^{-ik_yl} g_3 & g_1 g_3 =-i  g_2 \\
g_2\overline{g_1} =-i e^{i(k_x-k_y) l}  g_3 & g_2\overline{g_2} =
-e^{i(k_x+k_y)l} & g_2\overline{g_3} = ig_1 \\
g_3 \overline{g_1} =i e^{ik_x l} g_2 & g_3\overline{g_2} = ie^{ik_x l}g_1 &
g_3\overline{g_3}=1
\end{array}
\label{repeqs}
\end{equation}
At $\Gamma$ and $M$-points $g_2\overline{g_2} = -1$ and consequently there
are no one-dimensional representations.
The lowest dimension of irreducible representation of relations (\ref{repeqs}), realized by Pauli
matrices $g_i=\sigma_i$, is two. Thus, the 
eigenstates in the center and in the corner of 
the BZ are at least doubly
degenerate.

At points $X$ and $Y$ of Brillouin zone
$$
g_2^2\Psi(\br)=\Psi(\br),
$$
and there are two one-dimensional representations
$g_1=\pm 1$, $g_2=\pm i\lambda$, and $g_3=\lambda$, 
where $\lambda$ is an arbitrary complex 
parameter with unit norm $|\lambda|=1$.
The energy bands therefore are non-degenerate at $X$ and $Y$. These 
results are in complete agreement with the numerical solution 
obtained by Franz and Te\v sanovi\' c\cite{ft00} which is shown in Fig.
\ref{fig:theta0.25pim0.25pi}. 

To understand why the boundary  conditions 
with  $\theta_{1(2)}=\pm\pi/4$ are
favored by the plane wave expansion procedure, we recall that in this
method the eigenfunctions of ${\cal H}_{FT}$ are sought as 
\begin{equation}
\psi_{\bk}(\br)=\sum_{\bQ\in \Sigma_N}\begin{pmatrix}
u_Q^{\bk}\\
v_Q^{\bk}
\end{pmatrix}
e^{i(\bQ+\bk)\br}
\label{planewaves}
\end{equation}
where the set of reciprocal lattice vectors $\Sigma_N$ in the expansion form a
square grid $(2N+1)\times(2N+1)$:
\begin{equation}
\label{grid}
\Sigma_N = \left\{ \frac{2\pi}{l} (m,n):  -N\le m,n\le N \right\}~~,
\end{equation}
where $m$ and $n$ are integers. 
Here we make an assumption, well justified by numerical calculations, 
that the procedure produces convergent result as the number of plane waves in 
the expansion (\ref{planewaves}) is increased. Note that the choice of 
subset $\Sigma_N$ of reciprocal lattice vectors $\bQ$ is crucial, since
the matrix elements decrease only as $1/|\bQ|$ for large momenta, and 
if the matrix in reciprocal space was
truncated differently, the eigenstates could
end up corresponding to a {\em different}
self-adjoint extension.
After substitution into Hamiltonian ${\cal H}_{FT}$  the
coefficients $\Psi_{\bQ}^{\bk}=(u_{\bQ}^{\bk}, v_{\bQ}^{\bk})$ can be
found by diagonalization of the Fourier-transformed equations
\begin{equation}
H_0(\bk+\bQ)\Psi_{\bQ}^{\bk} + \sum'_{\bQ'\in \Sigma_N}\hat{V}(\bQ-\bQ')\; \Psi_{\bQ'}^{\bk} =
E\Psi_{\bQ}^{\bk}~,
\label{ftrecipr}
\end{equation}
where the summation is performed over all reciprocal vectors $\bQ'\in
\Sigma_N$ except $\bQ'=\bQ$, the zero field Hamiltonian $H_0(\bp)$ is
$$
H_0(\bp) = \sigma_3 p_x + \sigma_1 p_y~,
$$
and $(i\pi \hbar v_{BZ})^{-1}Q^2\hat{V}(\bQ) $ is given by
$$
\begin{pmatrix}
Q_y e^{-i\bQ\cdot\bR_A} & \frac{Q_x}{2} \left(e^{-i\bQ\cdot\bR_B}-e^{-i\bQ\cdot\bR_A}  \right)\\
\frac{Q_x}{2} \left(e^{-i\bQ\cdot\bR_B}-e^{-i\bQ\cdot\bR_A}  \right) & Q_y e^{-i\bQ\cdot\bR_B} \\
\end{pmatrix}
$$
Using identities
$\hatV(-\bQ)=\overline{\hatV(\bQ)}=-\sigma_2V(\bQ)\sigma_2$ and
$\sigma_2H_0\sigma_2=-H_0$
it is easy to verify that if $\Psi^{\bk}_{\bQ}$  is the 
eigenfunction of (\ref{ftrecipr}) with momentum $\bk$ and energy $E$ then
$$
\Psi'_{\bQ}=\sigma_2 \overline{\Psi_{\bQ}^{\bk}}, \quad \bQ\in\Sigma_N
$$
is also a solution of (\ref{ftrecipr}) with momentum $\bk$
and energy $-E$. In real space the new
wavefunction corresponds to 
$\Psi'(\br)=CP \Psi(\br)= \sigma_2\overline{\Psi(-\br)}$.
Similarly, from identity $\hatV(\bQ) = -e^{2\bQ\cdot \bR_A} \hatV(-\bQ)$ we
can obtain  another eigenstate
$$
\Psi''_{\bQ}=\Psi_{-\bQ}^{\bk} e^{2\bQ\cdot \bR_A}, \quad \bQ\in\Sigma_N
$$
with with energy $-E$ but inverted Bloch momentum $-\bk$. 
In real space the new
wavefunction is described by $I\Psi = \Psi(2\bR_A-\br)$. In fact using the
properties of matrix elements $\hatV$ it is straightforward to show that all
transformations of Table \ref{sympsi} from Appendix that 
do not involve factors $e^{-i\dphi}$ are
exact symmetries of the Hamiltonian (\ref{ftrecipr}) for any finite $N$.
According to our
assumption the procedure converges to a solution, 
which corresponds to some
self-adjoint extension described by certain  $\theta_{1(2)}$. 
By inspection, we find from Table
\ref{sympsi} from Appendix  that the only choice of $\theta_{1(2)}$ 
ensuring the above symmetries are satisfied
for arbitrary finite $N$ is $\theta_{1(2)}=\pm \pi/4$.

We conclude by demonstrating that this self-adjoint extension has a
single nodal point with zero energy at $\bk=0$. 
Since Hamiltonian ${\cal H}_{FT}$ does not
possess a small parameter, and both the scalar and vector potentials $\bv$
and $\ba$ are singular near vortices, it is desirable to base the discussion
on non-perturbative argument. Such an argument was suggested by
Vishwanath\cite{ashvini, ashvinii}, who made 
use of two properties of the energy
spectrum at  $\bk=0$. First, all states are doubly degenerate at $\bk=0$,
second, for each state $(\bk, E)$ there is another state $(\bk,-E)$.
He noted that the two properties are shared by the zero-field problem, which
has a pair of states at zero energy and symmetrically located pairs of
states at $E_n$ and $E_{-n}$, where $n$ is the band index. Thus, the number
of states is $2(2n_{max}+1)$, where $n_{max}$ is a heuristic parameter
denoting the number of bands with positive 
energy after fictitious ultraviolet cut-off is
introduced in order to make the spectrum bounded. 
According to Vishwanath, since the 
total number of states 
is preserved after the potentials $\bv$ and
$\ba$ are turned on, the pairs of 
degenerate states of ${\cal H}_{FT}$ must also be 
centered around zero. Otherwise the total
number states in the system would have been a $2(2n_{max})$.

The argument in the form just presented might seem ill-defined and lead to
inconsistent results. Indeed, if applied blindly, it would falsely suggest
that the spectrum should also be
gapless  at the corner of the Brillouin zone. 
Certainly, the number of states at
the center and the corner of BZ must be the same for any reasonable
high-energy cut-off and therefore should be equal to $2(2n_{max}+1)$. Just as
for the states at the center of BZ, the states at the corner are doubly
degenerate, and for each pair of states there is a
symmetric pair with the  opposite
energy. Thus, one might conclude, there must be a pair of states precisely at
zero energy not only for the ${\cal H}_{FT}$ but 
also for the zero-field problem!
Yet, the spectrum at the corner of BZ is gapped.

Clearly, the difficulties one encounters by applying the above argument
literally are related to the high-energy cut-off and have a technical
character. We show below that Vishwanath's 
construction\cite{foot_ashvin} can be made precise
in the framework of the self-adjoint extensions adopted in this article.
We have seen that the plane wave expansion corresponds to the extension
characterized by $\theta_{1(2)}=\pm\pi/4$ if 
the set of reciprocal wave vectors
of the basis forms a square grid (\ref{grid}) as $N\to\infty$.
At each finite $N$, there are $(2N+1)^2$ 
reciprocal lattice vectors $\bQ$ contained in the grid 
$\Sigma_N$ and each $\bQ$ is associated with two basis vectors $(1,0)$ and
$(0,1)$. Thus the dimensionality of
the basis is $2(2N+1)^2$. Since even for finite $N$ the solutions at $\bk=0$
appear as doublets at energy $E$ and another doublet at $-E$, the argument
of Vishwanath can be directly applied, and one of the
doublets should occupy $E=0$, otherwise the number of states would have been
a multiple of $4$ rather than $2(2N+1)^2$.

It is interesting to observe why the same argument cannot be applied to the
eigenfunctions of  ${\cal H}_{FT}$ at the corner of the Brillouin zone, and
the spectrum remains gapped.
Although at $\bk=\bpi$ the spectrum is degenerate in the limit of infinite
grid of reciprocal lattice vectors, for finite $N$ the degeneracy is only
approximate. As $N$ is increased, the low energy bands become nearly
degenerate, however at high enough energies the spectrum at $\bk=\bpi$
remains corrupted.
\psfrag{form_E}{$E$}
\psfrag{form_k}{$\bk$}
\psfrag{form_psi}{$\psi_{\bpi}$}
\psfrag{form_g2psi}{$\psi_{-\bpi}=g_2\psi_{\bpi}$}
\begin{figure}[tbh]
\centering
\includegraphics[width=\columnwidth,height=2in]{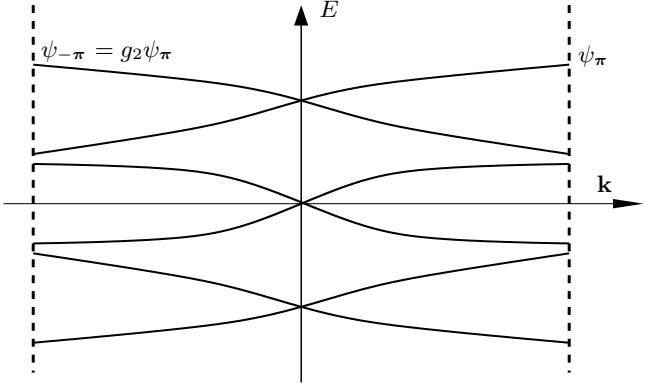}
\caption{\label{fig:ashvin} Schematic representation of the spectrum at
finite $N$. The small gaps between given two  neighboring bands at the corner of
Brillouin zone decrease to zero in the limit of infinite $N$.}
\end{figure}
Transformation $g_2=D(CPI)$, which ensured the degeneracy
of the spectrum at $\bk=0$ due to $g_2\overline{g_2}=-1$ even for finite $N$
does not have the same effect on the states at $\bk=\bpi$. 
Certainly, after applying $g_2$ to
an eigenstate $\psi_{\bpi}$ with momentum $\bpi$ and energy $E$, one still obtains  an
eigenstate $\psi_{-\bpi}=g_2\psi_{\bpi}$ with the same energy, but with momentum $-\bpi$. 
The situation is shown schematically in Fig.~\ref{fig:ashvin}.
Although for
infinite $N$ the points $\bpi$ and $-\bpi$ are equivalent, for 
any finite $N$ the wavefunctions describing the two states contain different
set of reciprocal lattice vectors, and therefore are distinct. In the
analysis of the degeneracy of the states the two values of $\bk$ should be
considered as independent.
Thus the argument on zero-modes cannot be applied at the corner of Brillouin
zone and for
$\theta_{1(2)}=\pm \pi/4$ there is a single Dirac node at $\Gamma$-point $\bk=0$.
\begin{figure*}
\includegraphics[width=\textwidth,clip]{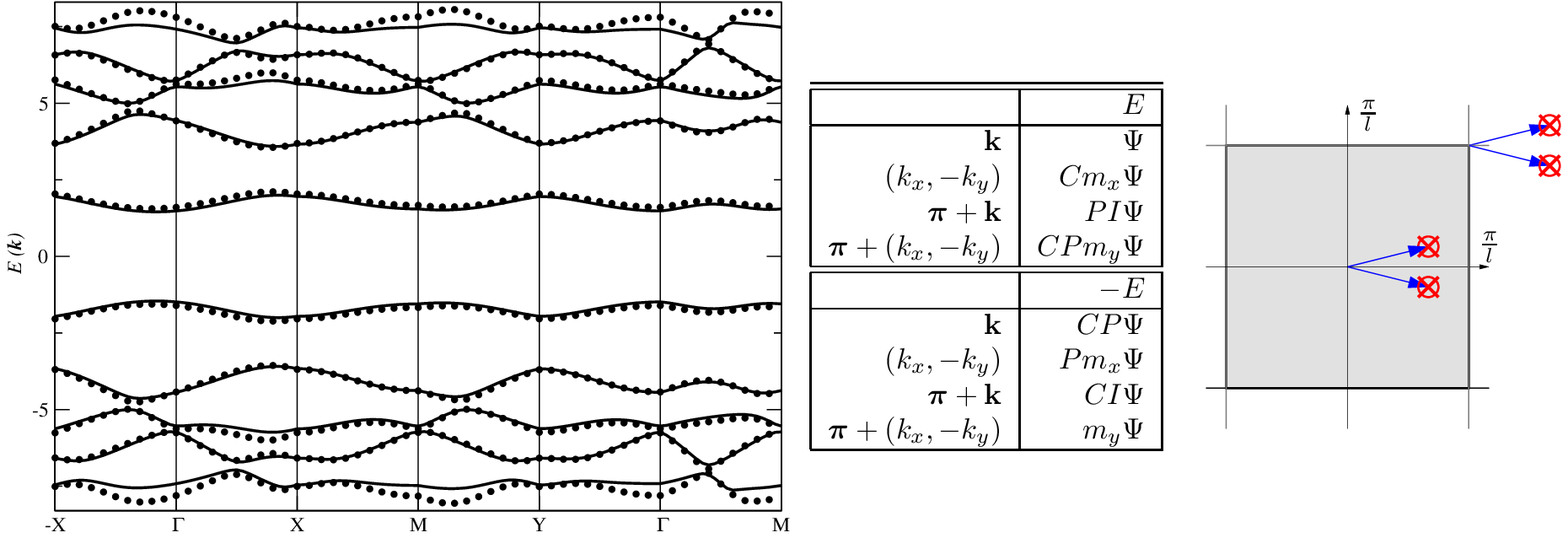}
\caption{\label{fig:theta0.0pi0.0pi}Left panel: the energy bands of linearized Hamiltonian for two
vortices per unit cell and boundary conditions 
$\theta_A=\theta_B=0$.  The spectrum obtained by OPW method is shown by
solid lines, the dots represent the spectrum of Hamiltonian (\ref{H_FTA})
with mass $M(\br)$ chosen as Gaussian $M_0\exp(-\br^2/\xi^2)$, centered
around each vortex with parameters $M_0=100/l$, $\xi=0.05l$. Small
discrepancies between the results obtained by two techniques for high energy
bands are due to finite size of the core size $\xi/l$ used in the numerical
calculation. Center panel: symmetry operations relating
eigenenergies at different $\bk$. Right panel: the first Brillouin zone
with equivalent $\bk$ vectors shown by  circles. For every state with energy
$E$ at momentum $\bk$ there is a state with the same momentum and energy
$-E$.}
\end{figure*}

\section{Numerical analysis}\label{num}
As we have shown, the straightforward expansion of  wavefunctions $\psi_{\bk}(\br)$ in the
plane wave basis 
$$
e^{i\bq\cdot\br}\begin{pmatrix}
1\\
0
\end{pmatrix}
\quad{\rm and}\quad
e^{i\bq\cdot\br}\begin{pmatrix}
0\\
1
\end{pmatrix}
$$
where $\bq=\bk+\bQ, \;\bQ\in\Sigma$ is problematic since the matrix elements of the Hamiltonian
decrease only as $1/\bQ$ because of the singular nature of the
perturbation potentials at vortex locations. Consequently the eigenvalues of
the  linearized Hamiltonian
depend on the way the matrix $\langle\bq'|H|\bq\rangle$ is truncated and 
consistent ultraviolet cut-offs are parameterized by parameters $\theta$
specifying the boundary conditions at
each vortex. For example, the
symmetric choice of the set $\bQ\in \Sigma_N$  results in a choice
$\theta_{1(2)}=\pm\pi/4$.
To enforce the desired boundary conditions we use a procedure similar to
orthogonalized plane waves method used in the standard band theory. Instead
of expanding the wavefunctions $\psi(r)$ in plane wave basis we define
\begin{multline}
\tilde{\chi}(\bR_i+\br) = \frac{e^{-r^2/a^2}}{(2\pi)^{3/4}\sqrt{ar}}
\left(
\sin\theta e^{\frac{i}{2}\cos\phi}
\begin{pmatrix}
e^{-i\phi}-i\\
ie^{-i\phi}-1
\end{pmatrix}
-\right.\\
\left.\cos\theta e^{-\frac{i}{2}\cos\phi}
\begin{pmatrix}
e^{i\phi}+i\\
ie^{-i\phi}+1
\end{pmatrix}
\right)
\end{multline}
if $\bR_i$ is an $A$ vortex, and
\begin{multline}
\tilde{\chi}(\bR_i+\br) = \frac{e^{-r^2/a^2}}{(2\pi)^{3/4}\sqrt{ar}}
\left(
\sin\theta e^{\frac{i}{2}\cos\phi}
\begin{pmatrix}
1-ie^{i\phi}\\
i-e^{i\phi}
\end{pmatrix}
-\right.\\
\left.\cos\theta e^{-\frac{i}{2}\cos\phi}
\begin{pmatrix}
1+ie^{i\phi}\\
i+e^{-i\phi}
\end{pmatrix}
\right).
\end{multline}
if $\bR_i$ belongs to class $B$. The wavefunctions that  have the same
asymptotics at the vortices and Bloch periodicity can easily then be
constructed as
$$
\chi_{\bq}(\br) = \sum_{\bR} e^{i\bq \bR} \chi(\br+\bR)
$$
The wavefunctions are orthogonal to the singular parts of asymptotics
(\ref{AsymptA}). The appropriate basis for expansion of the eigenfunctions
of the Hamiltonian ${\cal H}_{FT}$ with boundary conditions $\{\theta_i\}$ is
$$
e^{i\bq\cdot\br}
\begin{pmatrix}
1\\
0
\end{pmatrix}
-\sum_i\alpha^{i}_{\bq}\chi^i(\br)
\quad{\rm and}\quad
e^{i\bq\cdot\br}
\begin{pmatrix}
0\\
1
\end{pmatrix}
-\sum_i\beta^{i}_{\bq}\chi^i(\br)
$$
where cutoff $a$  is introduced for numerical convenience. The 
coefficients $\alpha^i_{\bq}$ and $\beta^i_{\bq}$ are chosen so that
the basis functions are orthogonal to $\chi^i(\br)$. In the limit  $a\to 0$
the procedure ensures that the modes with asymptotics orthogonal to those
given by $\theta$ are projected out. Because of the exponential
factors introduced in $\chi(\br)$ the largest reciprocal vector $\bQ$ should
be of order $1/a$ or larger, so that there is appreciable region around a
vortex  $1/Q_{max}<r<a$ where the wavefunctions have the desired
$1/\sqrt{r}$ asymptotics but still are not affected by the exponential factors.
We typically used $a=0.05 l$ and increased the number of basis elements
until the convergence is achieved. Further decrease of the damping parameter 
$a$ affects the results very weakly. 
Examples of the energy bands obtained by this method
for two vortices per unit cell for different boundary conditions  are
shown in Figs.
\ref{fig:theta0.25pim0.25pi} and
\ref{fig:theta0.0pi0.0pi}.
The first figure
shows comparison of the energy spectra for the case $\theta_1=-\pi/4$,
$\theta_2=\pi/4$
calculated by using the technique just described and
straightforward plane wave expansion.
The two are identical within numerical precision, confirming yet again that
simple plane wave expansion generates a solution described by $\theta_{1(2)}=\pm\pi/4$.

Quite a different result is obtained for boundary conditions
$\theta_A=0$, $\theta_B=0$. The spectrum in this case is gapped. The
symmetry analysis can be applied to describe the spectrum just as in the
previous section. Relevant symmetry transformations and the star of
equivalent symmetry points are shown in Fig.~\ref{fig:theta0.0pi0.0pi}. The features of the
numerical results fully agree with the results of numerical calculation: for example the
energy bands are  symmetric under transformation $(E,\bk)\to(-E,\bk)$,
and the segments $XM$ and $YG$ are equivalent.

As was pointed out, for  general boundary condition even the symmetry
$(E,\bk)\to(-E,\bk)$ of the single node energy spectrum is absent.
\section{Relation of the singular linearized Hamiltonian to the
non-linearized, regular problem}\label{relation}



As explained in previous sections, 
the spectrum  of linearized Hamiltonian, in which
vortices appear as point-like defects, depends on the boundary conditions
imposed at vortex locations. These boundary condition are in turn determined
by the self-adjoint extension of the linearized model.
In principle, in order
to establish which particular self-adjoint extension should be
used, one faces solving the 
original, fully self-consistent BdG equations, a project beyond the
scope of the present paper. 

Still, in the
simplest case of a $d$-wave superconductor 
with no additional forms of ordering in the 
interior of vortex cores, the physics of nodal quasiparticles should
be adequately described by the Hamiltonian of Eq. (\ref{bdg}) and we should
be able to say something definite about the meaning of different self-adjoint
extensions. A useful perspective on various 
choices of boundary conditions, i.e.
various choices of the parameter $\theta$, can be gained by noticing 
that the main feature of the self-consistent solution is to
suppress the quasiparticle wavefunctions inside vortex cores.
Furthermore, such fully self-consistent solution does not require any
additional boundary conditions at vortex locations, since
non-linear terms, which
grow strongly near vortices, and the
self-consistency condition conspire 
to regularize the wavefunctions inside the core.

How can such behavior be emulated 
on the level of a Dirac Hamiltonian {\em without}
explicitly enforcing the boundary conditions?
Unlike particles described by a Shcr\"odinger Hamiltonian, 
Dirac fermions cannot be prevented from penetrating vortex cores by
a strong scalar potential barrier at vortex locations -- if such a barrier is
imposed it leads to the Klein paradox\cite{berry}.
The correct procedure which ensures suppression of the Dirac spinor amplitude
in a particular region of space
requires that the mass of the particle be
treated as a function of position $M(\br)$,
so that in the prohibited region such 
mass becomes very large\cite{berry}. If the mass of
the particle in the interior of forbidden region is set to 
$M_0$, then the spinor wavefunctions experience an exponential suppression
in this same region, with a penetration length $\sim 1/M_0$.
In order to regularize the FT Hamiltonian by requiring suppression of the
nodal quasiparticle 
wavefunctions inside vortex cores, we are thus
led to introduce a short-ranged
mass-like potential $\sigma_2 M(\br)$  that vanishes at distances larger
than the core size $\sim \xi$. Inside the core, 
the absolute value of  mass $M_0$ should be
chosen to be much larger  than $1/\xi$. 

Even in the zero-field problem such mass term breaks time-reversal symmetry:
$\sigma_2 M(\br)$ changes sign under 
the time-reversal operation, and one might expect that
in such terms in general lead to opening of a gap in the 
spectrum. The detailed
analysis shows that FT equations augmented by mass potentials
\begin{equation}
\label{H_FTA}
\frac1{v_F}{\cal H'}_{FT} =  (p_x+a_x)\sigma_3 
+\alpha_{\Delta} \sigma_1(p_y+a_y) + v_x +\sigma_2 M(\br)
\end{equation}
are generally {\em not} invariant under symmetry operations listed in 
Table~\ref{sympsi} from Appendix. The transformation properties of the mass term itself
$M(\br)$ are shown in Table~\ref{masstransform}.
\begin{table}
\begin{tabular}{|c|c|c|c|}
\hline
$1$ & $I$ & $P$ & $PI$\\ 
\hline
$M(\br)$  &
$-M(I\br)$&
$-M(P\br)$&
$M(s\br)$ \\
\hline
\hline
$C$  & $CI$ & $CP$ & $CPI$\\
\hline
$-M(\br)$ &
$M(I\br)$ &
$M(P\br)$ &
$-M(s\br)$\\
\hline
\end{tabular}
\caption{\label{masstransform} Transformation of the
position-dependent mass term $M(\br)$ under operations commuting
with  $H_{FT}$. Only for $M(\br)$ of special symmetries the new
Hamiltonian $H'_{FT} = H_{FT}+\sigma_2 v_FM(\br)$ remains invariant upon the
action of these operations.
}
\end{table}
Clearly, {\em no} choice of nonzero mass term $M(\br)$ is invariant under
{\em all} symmetry operations, just as no choice of $\theta_i$'s
preserves {\em all} the symmetries of 
$H_{FT}$. For $M(\br)$ possessing 
certain specific symmetries, some of the
transformations do commute with the Hamiltonian. 
For example, if the mass does
not change under inversion about a 
vortex and has the same sign for all
vortices:
$$
M(I\br)=M(\br)=M(s\br),
$$
then the transformations commuting with 
the Hamiltonian coincide with those listed in
Fig.~\ref{fig:theta0.0pi0.0pi}. Thus the 
Hamiltonian with such mass terms is {\em equivalent}
to the original FT Hamiltonian with $\theta_1=\theta_2=0$ and with 
mass equal to zero. The spectrum of ${\cal H}'_{FT}$ 
calculated by the standard plane wave expansion
is shown in Fig.~\ref{fig:theta0.0pi0.0pi} and agrees with the numerical
results based on the OPW method.
Similarly, a straightforward calculation 
shows that the choice of  
$$M(\br)=M(I\br)=-M(P\br)$$
corresponds to $\theta_1=0;~\theta_2=\pi/2$, while
$$M(\br)=-M(I\br)=-M(P\br)$$ 
describes 
$\theta_1=-\theta_2=\pi/4$. 
The latter choice requires that
$M(\br)$ has a $p$-wave symmetry and vanishes at least along two directions
around each vortex. This choice also illustrates the following important
point: while the mass term always breaks time-reversal
symmetry, in the $\theta_1=-\theta_2=\pi/4$ case it does so
only {\em locally} while leaving this 
symmetry intact on {\em average} since $\int d^2r M(\br) =0$. Here
the integration is over the small region within which $M(\br)$ differs 
appreciably from zero around a {\em single} vortex. 

We will conclude this section with two observations. First, if the overall
amplitude of the mass term is increased continuously, then the gap in the
spectrum will also increase, and it is not {\em a priori} clear that in the
limit of extremely small magnetic field the magnitude of the gap induced by
the mass term is proportional to $1/l$ rather than $1/l^2$. To answer that question we note that
if the absolute value of the mass $M(\br)$ inside vortex core 
is denoted as $M_0$, then the quasiparticles  penetrate into
the vortex core to  distances of order  $1/M_0$. Since the distance
should be much smaller than the size of the vortex core  $\xi$, 
 we find that $M_0\gg l/\xi$. The scaling limit,
therefore is obtained as a limit of infinite $M_0$.

Our analysis does not allow us to uniquely determine which
self-adjoint extension (specified by $\theta_{1,2}$ or by  symmetry of the
mass terms) is realized in cuprates,
as the answer depends  significantly on short range physics and
can be determined only through the detailed  analysis of the spectrum and the
wavefunctions of self-consistent solution. However, given the extension of
the Hamiltonian describing quasiparticles near node $1$, the symmetry properties of the
full BdG Hamiltonian determine unambiguously the self-adjoint
extensions of  linearized Hamiltonian near nodes $\bar{1}$, $2$ and $\bar{2}$.

Symmetry operations of the linearized 
Hamiltonian listed in (Table \ref{sympsi} in the Appendix)
also commute with the full BdG Hamiltonian
${\cal H}$ defined by (\ref{H_FT_nonlin}), where now these symmetry
operations are applied to the full wavefunction $\Psi$. One should 
note, however, that these operation
have entirely different meaning as ${\cal H}$ describes simultaneously all four
nodes and most of the operations  relate the eigenstates belonging to
different nodes. For example, the action of the inversion operator on an
eigenfunction $\Psi(\br)$ of ${\cal H}$ will now
be understood as $\Psi(-\br)$:
$$
\Psi(\br)\approx e^{ik_F x}\psi(\br) \to  e^{-ik_F x}\psi(-\br)~;
$$
the reader should compare this to the ``inversion'' operation of 
Eq. (\ref{blah}).
Thus, it transforms the state near node $1$ into 
a state with Fourier components
localized in the vicinity of $\bar{1}$.
Similarly, symmetry operations  involving
reflections $m_x$ and $m_y$ applied to quasiparticle wavefunctions with
Fourier components localized near node $1$, generate new wavefunctions near
nodes $2$ and $\bar{2}$. 

\begin{figure}[tbh]
\centering
\includegraphics[width=\columnwidth]{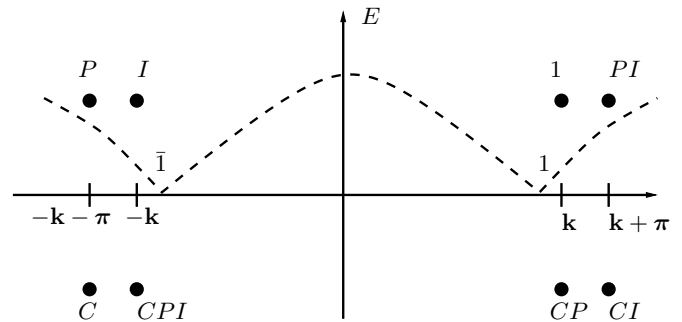}
\caption{\label{fig:fullsymm}Symmetry of the spectrum for
the full BdG Hamiltonian ${\cal H}$.}
\end{figure}

For simplicity, we will focus on transformations generated by $P$,
$I$, and $C$, which relate  nodes $1$ and $\bar{1}$ only.
Acting on a wavefunction $\psi(\br)$ with crystal momentum $\bk$ and energy $E$, they generate
seven new states shown schematically in Fig. \ref{fig:fullsymm}.
\psfrag{f_la_E}{$E$}
\psfrag{f_la_node1}{$1$}
\psfrag{f_la_node1bar}{$\bar{1}$}
\psfrag{f_lak}{$\bk$}
\psfrag{f_lakpi}{$\bk+\bpi$}
\psfrag{f_lak1}{$PI$}
\psfrag{f_lak2}{$1$}
\psfrag{f_lak3}{$CP$}
\psfrag{f_lak4}{$CI$}
\psfrag{f_lamk}{$-\bk$}
\psfrag{f_lamkpi}{$-\bk-\bpi$}
\psfrag{f_lamk1}{$I$}
\psfrag{f_lamk2}{$P$}
\psfrag{f_lamk3}{$C$}
\psfrag{f_lamk4}{$CPI$}

Note that transformations $PI$, $CP$, and $CI$ relate wavefunctions near the
same node. They ensure
that if $\Psi(\br)$ is an eigenstate of
Hamiltonian (\ref{bdg}) with energy $E$  and momentum $\bk$ then
the spectrum of ${\cal H}$ also has states characterized by  $(\bk,-E)$,
$(\bk+\bpi, E)$, and   $(\bk+\bpi, -E)$.

One can easily verify that there are only two self-adjoint extensions
compatible with $PI$, $CI$ and $CP$:  $\theta_1=\theta_2=0$ and $\theta_1=\theta_2=\pi/2$.
The two choices are essentially equivalent being related to each other by
reflection $m_x$. The resulting single-node energy spectrum is shown in
(\ref{fig:theta0.0pi0.0pi}). It is characterized by strongly 
dispersive bands at
low energies, except for the lowest gapped  narrow bands at energies $\approx
\pm 2\hbar v_F/l$. 

All other boundary conditions break one of the three symmetries.
In particular, if the spectrum is gapless at $\bk^F_1$ then it should
 also be gapless at momentum $\bk^F_1+\bpi$. This is clearly not the case
for the choice $\theta_1=\pi/4$, $\theta_2=-\pi/4$ shown in Fig.
\ref{fig:theta0.25pim0.25pi}. To restore the full 
symmetry of the non-linearized BdG
Hamiltonian (\ref{bdg}), one has to demand that the 
linearized version is a
{\em union} of two independent self-adjoint 
extensions: $\theta_{1}=-\theta_2=\pi/4$ and
$\theta_{1}=-\theta_2=-\pi/4$. Such union is a combination of the 
solution found in Ref. \onlinecite{ft00} and its companion solution with the
A and B labels interchanged (DOS, of course, remains unaffected apart
from a mere factor of two).
Consequently, the single-node linearization corresponds to
{\em two independent} problems described by the same Hamiltonian operator, but
with different boundary conditions near vortices. Under $PI$ operation, the Hilbert spaces of the two Hamiltonians are interchanged and the full
symmetry is thereby restored. Observe that this is a deeply non-perturbative
result: such a union describes {\em eight} two-component Dirac
fermions and therefore 16 zero energy states at BdG nodes, in contrast
to only four massless Dirac fermions of the $H=0$ case.
Recent exact symmetry arguments on a tight-binding lattice\cite{index},
based on a form of the index theorem, strongly support the identification
of the above union as the proper self-adjoint extension 
of the continuum linearized problem with gapless spectrum.

\section{Linearization of $d_{x^2-y^2}$ Hamiltonian}
\label{orient}
In the companion paper \cite{tight},
we consider properties of a tight-binding lattice
$d$-wave superconductor in the mixed state. The simplest representative of
such systems is characterized by the $d_{x^2-y^2}$ symmetry of
the order parameter rather
than the $d_{xy}$ case considered in previous sections. 
In order to directly compare
the linearized model and the tight-binding lattice model, we use this
section to consider the linearized 
Hamiltonian with the ``clover'' of the $d$-wave
gap function
rotated by 45 degrees with respect to the vortex lattice. After following
the standard steps\cite{vmft,tight}, we
find that the linearized Hamiltonian in this case is given by
\begin{equation}
H_{x^2-y^2}^{lin} = v_F \frac{\Pi_x+\Pi_y}{\sqrt{2}}\sigma_3
+v_{\Delta}\frac{\Pi_y-\Pi_x}{\sqrt{2}}\sigma_1
+v_F \frac{v_x+v_y}{\sqrt{2}}
\label{H_TB_lin}
\end{equation}
where  $\Pi_i = p_i+a_i$ is the generalized momentum.
In the rest of the section, we consider the properties of the above
Hamiltonian for precisely the same orientation and position of the vortex
lattice as before, with the unit cell shown in Fig. \ref{fig:unitcell}.
Therefore, the potentials $\bv$ and $\ba$ remain exactly the same, and the only
difference from the original Hamiltonian ${\cal H}_{FT}$ is the direction of
the $d$-wave nodes relative to the vortex lattice.

As before, we will confine ourselves to the isotropic case 
$v_F=v_{\Delta}$.  Note that in this case, if one 
neglects the superfluid velocity terms and
retains only the vector potential $\ba$,   Hamiltonians
$H_{x^2-y^2}^{lin}$ and $H_{FT}\equiv H^{lin}_{xy}$ are related by a unitary
SU(2) transformation $U=\exp(i\sigma_y \pi/8 )$. Therefore, both are
expected to have the same density of states. Moreover, since this unitary
transformation is global, the dispersions $E_n(\bk)$ 
of the two cases will also be identical.

The above invariance is a consequence of the combination of
our Dirac particles 
effectively being at
$\bk=0$ and the axially-symmetric character of the free
Dirac problem. Since there is no preferred direction for the 
free Dirac problem, different orientation of the vortex lattice cannot
change the spectrum. This invariance, of course, 
holds for any linearized Hamiltonian 
with an arbitrary  direction of the  $d$-wave ``clover''. The unitary
transformation in this general case, relating the Hamiltonian
to $H_{FT}$ is
$U=\exp(i\sigma_y\alpha/2)$, where $\alpha$ is the direction of the node
relative to $x$-axis.

\begin{figure}[tbh]
\centering
\includegraphics[width=\columnwidth, clip]{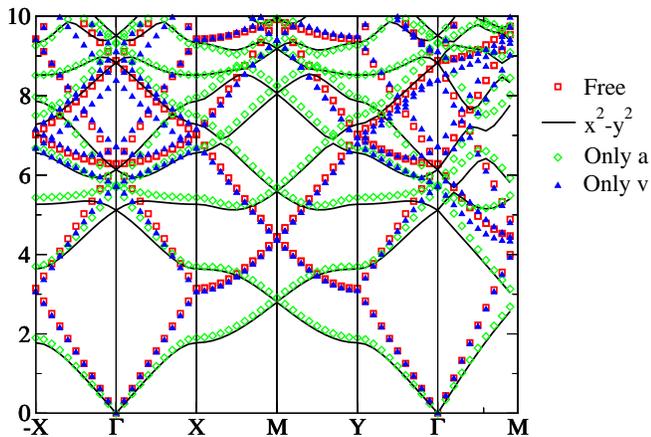}
\caption{\label{fig:spectraall} The quasiparticle spectrum of the mixed state
obtained by an expansion of the wavefunctions in
the plane wave basis. The dispersion of the zero magnetic field
problem is shown in red squares,  the dispersion of a $d_{x^2-y^2}$
superconductor in the presence of a vortex lattice is shown by a solid black
line. The green diamonds (blue triangles) 
correspond to the artificial  Hamiltonian, which is
obtained by setting the scalar potential $\bv$ (vector potential
$\ba$) in $H_{FT}$ to zero. Note that the role of the scalar potential at low energies
is small, and that makes the spectrum of  $H^{lin}_{x^2-y^2}$ rather
similar to $H^{lin}_{xy}= H_{FT}$ shown in  Fig. \ref{fig:theta0.25pim0.25pi}.}
\end{figure}

Of course, this invariance is violated by the superfluid
velocity terms and finite anisotropy. In the latter case, there {\em is} a
preferred direction since the free Dirac dispersion defines an elongated cone.
Similarly, the terms containing the superfluid velocity $\bv$ carry the
information about a particular direction of the node, where the 
linearization is
performed, and therefore break the rotational symmetry, even when $\ba$ is
set to zero. Numerically, the effects of $\bv$ turn out to be rather
small at the lowest band energies, and become progressively more important
for higher bands. Fig. \ref{fig:spectraall} shows the spectrum of
$d_{x^2-y^2}$ superconductor obtained numerically using  a simple expansion
in plane waves. For the lowest bands, it is very similar to the spectrum of
$d_{xy}$-superconductor discussed in detail in the earlier sections
(see Fig. \ref{fig:theta0.25pim0.25pi}). In both
 $d_{xy}$ are $d_{x^2-y^2}$ cases, for the lowest band the result is very
close  to that of  the reduced Hamiltonian, which contains only the
vector potential $\ba$ -- the effect of the scalar potential becomes
pronounced only at higher energies. The spectrum of the Hamiltonian with
only  the scalar potential $\bv$ included and the  vector potential set
to zero by hand, is quite different -- at low energies it is a tiny
perturbation of the free Dirac dispersion. This is in stark contrast with
the result of the so-called ``Volovik approximation.'' \cite{volovik}

The allowed class of the boundary conditions near vortices for
$H^{lin}_{x^2-y^2}$ can be obtained
directly from the boundary conditions of the $d_{xy}$ case (\ref{bcA}) and
(\ref{bcB}) by noting that in the coordinate system $(x',y')$  rotated  by
45 degrees with respect to $(x,y)$,
$H^{lin}_{x^2-y^2}$ takes on the form identical to  $H_{FT}$,
except for the vortex lattice being rotated by $\pi/4$ 
relative to what it used
to be in the $d_{xy}$ case. Still, when deriving the asymptotics of the
wavefunctions near a vortex, only the singular terms due to that single
vortex contribute -- different geometry of the lattice
therefore does not directly affect the boundary
conditions. All we need to do is to return to the original
coordinate system by performing the rotation $\phi\to \phi-\pi/4$  in
(\ref{bcA}) and (\ref{bcB}).

The symmetry properties of the Hamiltonian $H^{lin}_{x^2-y^2}$ can be
derived rather straightforwardly, since $\bv$ and $\ba$ are 
exactly the same as before, and we
can use their properties from 
Table \ref{symvAB}. The symmetry properties of the
Hamiltonian, which do not involve mirror symmetries, remain exactly the
same. Namely, transformations $I, P, C$ and their combinations can be
literally taken from  Table \ref{sympsi}. The transformations involving
mirror symmetries are modified, however. Moreover, in addition to the mirror
planes along the $x$ and $y$ directions, now a new mirror symmetry plane
appears along the diagonal of the unit cell. The appearance of these new
elements of symmetry is important for a precise form of the spectrum. 
It is easy to see from Figs. \ref{fig:theta0.25pim0.25pi} and  \ref{fig:spectraall} that,
although the spectra of the $d_{xy}$
and $d_{x^2-y^2}$ are quite close,  the latter is more symmetric
due to the additional symmetry $x\leftrightarrow y$.

The main conclusions, nevertheless, remain the same. For example, 
the symmetry of
the Hamiltonian operator requires that the spectrum is
symmetric under translations in momentum space by $(\pi/l,\pi/l)$. In other
words, if there is a node at $\bk=0$, it should have been replicated at
the corner of the unit cell. Clearly, the dispersion  
of Fig. \ref{fig:spectraall}
has only one gapless point at $\bk=0$. The reason, just as before, is the
necessity of imposing the boundary conditions which again violate the
symmetries of what is formally the Hamiltonian operator.

\section{CONCLUSIONS}
The main results of this paper can be summarized as follows:
the linearized BdG Hamiltonian describing nodal fermions in the mixed state of
a $d_{x^2-y^2}$ superconductor has to be
complemented by a set of boundary 
conditions (\ref{bcA},\ref{bcB}) specifying the 
behavior of quasiparticle wavefunctions near vortices 
in order for the problem to be
mathematically fully defined. The boundary conditions 
contain a single parameter $\theta\in[0,\pi)$
for each vortex which cannot be 
found from within the linearized
theory itself, but should be determined from the 
full non-linearized calculation. 
All the physics beyond linearization, such
as the intervortex scattering and interference effects, particle-hole
asymmetry, high energy and curvature terms, etc., is implicitly
reflected in the effective linearized FT
Hamiltonian {\em only} through these boundary conditions.
Consequently, the linearized FT Hamiltonian actually
describes a {\em family} of
distinct self-adjoint extensions reflecting a variety of high-energy 
processes that might be taking place inside the vortex cores
(magnetism, Mott insulator, charge density-wave, etc.).
Such multitude of all possible short-range physics behaviors can be
classified according to the set of $\theta$ parameters, which govern
different ``universality'' classes of the 
nodal fermion quantum ``criticality.''
The conventional Simon-Lee
scaling function must be generalized to include the explicit dependence
on $\theta$'s, as explained in the text. 
Once the boundary conditions are properly 
imposed according to (\ref{bcA},\ref{bcB}),
the solutions of the FT equations are fully determined and are independent of
the assignment of A or B vortices, thereby restoring their invariance under
arbitrary singular gauge transformations.

In earlier work\cite{ft00} the boundary conditions 
were not explicitly enforced
but instead were ``spontaneously'' selected 
by the procedure used to
diagonalize the Hamiltonian. Such an expansion in the plane wave
basis results in $\theta_{1(2)}=\pm\pi/4 (\mp\pi/4)$, which was indeed
found here to be the self-adjoint extension appropriate to the nodal 
gapless behavior of a $d$-wave superconductor in the limit
of low magnetic fields.
In general, however, the boundary conditions $\theta_{1(2)}$ 
should be found from the full
self-consistent BdG Hamiltonian  
and serve as an external input to the linearized theory
describing the bulk quasiparticle states.

The authors thank Dr. O. Vafek for useful discussions and Profs. 
M. R. Zirnbauer and A. Altland for helpful 
correspondence and for sharing with us their unpublished results.
This work was supported in part by the NSF grants DMR-0094981
and DMR-0531159.

\appendix

\section{Superfluid velocity and the phase of superconducting order parameter}

\label{sup}
\subsection{Simple properties}

\begin{figure}[tbh]
\centering
\includegraphics[width=\columnwidth]{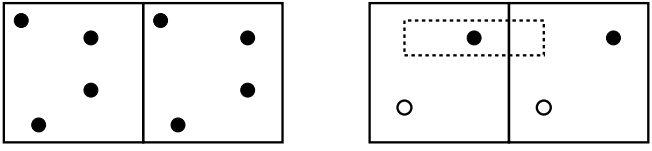}
\caption{\label{fig:proof} Left panel: An example of a vortex lattice for the
case of N=4  vortices per unit cell. 
Right panel: Two vortices per unit cell: integral
(\ref{contint}) calculated along the contour shown as a  
dashed line equals $2\pi$.}
\end{figure}

For the reference,  we summarize the properties of superfluid velocity and the
phase of the order parameter used in the text. Consider a periodic lattice of
vortices with a basis, with $N$ vortices located at $\bR_i$ within a square
unit cell of size $l\times l$ as shown in Fig. \ref{fig:proof}.
%
%
The superfluid velocity is  defined as
\begin{equation}
\label{vdef}
m\bv=\frac12\hbar \nabla \phi -\frac{e}{c} \bA~~,
\end{equation}
where  $\phi(\br)$ is the phase of the order parameter.
The latter can be determined from the requirement that it
acquires $2\pi$ as $\br$ encircles a vortex:
\begin{equation}
\label{deltacurl}
\nabla\times\nabla \phi=2\pi\hat{z}\sum_{\bR}{\delta(\br-\bR)}~~,
\end{equation}
where $\bR$ denotes the position of the vortices.
Since the superfluid velocity typically enters in combination $m\bv$
we will often set $m=1$, for
compactness.


Using the Fourier transforms, the superfluid velocity
can be written \cite{tinkham} as
\begin{equation}
\bv (\br)= i\pi \hbar
\int\frac{d^2 k}{(2\pi)^2}\frac{\bk\times\hat{z}}{\lambda^{-2}+k^2}
e^{i\bk\cdot\br}\sum_{\bR} e^{-i\bk\cdot \bR}~~,
\label{vfour}
\end{equation}
where $\lambda$ is the London penetration length. Let us write the position of a
vortex as $\bR=\bR_i+\btau$, where  index $i=1{\ldots} N$ labels
the vortices inside an arbitrarily chosen reference unit cell, and $\btau=l(N_x\hat{x}+N_y \hat{y})$ with
integer $N_x$, $N_y$ denotes lattice translation vectors.
Then, after summation over $\btau$,  the superfluid velocity can be expressed  as
\begin{equation}
\bv(\br) = \frac{i \pi \hbar}{l^2}\sum_{\bQ} \frac{(Q_y,
-Q_x)}{Q^2+\lambda^{-2}}e^{i\bQ\br}\sum_{i=1}^N e^{-i\bQ \bR_i}~~,
\label{superv1}
\end{equation}
where $\bQ=2\pi(\hat{x} N_x +\hat{y}N_y)/l$ with integer $N_x$, $N_y$ are the reciprocal lattice vectors.
Note that when $\lambda\gg l$, one can neglect  $\lambda^{-2}$ in the
denominator in all terms of the sum except for $\bQ={\bf 0}$, and the simplified result reads:
\begin{equation}
\bv(\br) = \frac{i \pi \hbar}{l^2}\sum_{\bQ\neq 0} \frac{(Q_y,
-Q_x)}{Q^2}e^{i\bQ\br}\sum_{i=1}^N e^{-i\bQ \bR_i}~~.
\label{superv2}
\end{equation}
This simplification  corresponds to replacing the
magnetic field $H(\br)$ by its spatial average, as can be verified directly by
comparing $\nabla\times \bv$ from (\ref{superv1}) and (\ref{vdef}). 
In the rest of the appendix and in the main text, the limit $\lambda\gg l$ is
always implied, and $A(\br)$ denotes the vector potential corresponding
to the uniform applied magnetic field, without including small  corrections
due to the vortex lattice, which are smaller by a factor of $l^2/\lambda^2$.

The superfluid velocity $\bv$ can be  expressed in 
a closed form through Weierstrass elliptic zeta
function (see also Ref. \onlinecite{phiaphib}). 
To derive this expression, let us start from  (\ref{deltacurl}), and
write the phase $\phi$ as 
$$
\phi(\br)=\phi_0(\br)+\sum_{\bR} \arctan \frac{y-Y}{x-X}~~,
$$
where  $\phi_0(\br)$ is a continuous function to be determined later
and the second term is 
a sum of the polar angles describing $\br$ with respect to the vortex location $\bR$.
After denoting $z=x+iy$ and substituting $\phi$ into (\ref{vdef}) we obtain
\begin{multline}
\label{zetadef}
v_y + i v_x =
\sum_{i=1}^N\frac{\hbar}2\sum_{\tau}\frac1{z-Z_i-\tau} - \frac{e}{c}(A_y+iA_x)\\+
\frac12\left(\nabla_y\phi_0+i\nabla_x\phi_0\right)~~,
\end{multline}
where the lattice vectors are $\tau=\tau_x+i\tau_y$ and the location of
the vortices within the reference unit cell are now encoded by $Z_i=X_i+iY_i$.
The sum over the lattice vectors $\tau$ in the right hand side differs from the definition of Weierstrass
zeta-function
\begin{equation}
\label{zetadef1}
\zeta(z) = \frac1z+\sum_{\tau\neq 0 }\left(\frac1{z-\tau} +\frac1{\tau}
+\frac{z}{\tau^2}\right)
\end{equation}
only by a presence of the constant  and linear in $z$ terms, which
are required to ensure the absolute convergence of the sum (\ref{zetadef1}).
These terms can be absorbed into the ``smooth'' part of the phase by defining  
$\phi_0'=\phi_0+C_0 z+\frac12 C_1z^2$, where $C_{1,2}$ are appropriately chosen
constants. Thus, the superfluid velocity satisfies the following equation:
\begin{equation}
\label{zetadef1A}
v_y+iv_x=
\frac{\hbar}2\sum_{i=1}^{N}
\left(\zeta(z-Z_i)-\frac{\pi\hbar\overline{z}}{l^2}\right)+
\frac{\nabla_y\phi_0'+i\nabla_x\phi_0'}2~.
\end{equation}
Here, for convenience we chose the symmetric gauge:
$$
\frac{e}{c}\bA = \frac{\pi N}{2l^2} (-y,x)~~.
$$
Note now that $w(x,y)=\nabla_y \phi_0' + i \nabla_x \phi_0'$  satisfies both
Cauchy-Riemann conditions: $\partial_x ({\rm Re} w) =\partial_y ({\rm Im} w)$ is
satisfied automatically, whereas  the second condition 
 $\partial_y ({\rm Re} w) =-\partial_x ({\rm Im} w)$ is satisfied due to
the requirement ${\rm  div}\; \bv =0$ and (\ref{zetadef1A}). Since $\phi_0$ (and
consequently $\phi_0'$) 
was chosen as a  smooth part of the phase, it is therefore a finite
analytic function of $z$ in the entire complex plane, and by Liouville theorem must be equal
to a constant. The remaining constant is fixed by the requirement for the spatial average
of $\bv$ to vanish (cf. \ref{superv2}), and the final result, which is equivalent to
(\ref{superv2}), reads 
\begin{equation}
\label{zetadef2}
v_y+iv_x=
\frac{\hbar}2\sum_{i=1}^{N}
\left(\zeta(z-Z_i)-\pi\hbar\frac{\overline{z-Z_i}}{l^2}\right).
\end{equation}
Note that the superfluid velocity $\bv(\br)$ is periodic in space with the
same unit cell as the unit cell of the vortex lattice. 


\subsection{The phase of the order parameter}
Unlike the superfluid velocity components, neither the 
phase of the order parameter
$\phi(\br)$, nor $\exp(i\phi)$, and not even $\nabla \phi$ are periodic,
regardless of the gauge used. 
Since the periodicity of  $\exp(i\phi)$ would have required
 that  $\nabla\phi$ is itself
periodic, it is sufficient to prove the statement for $\nabla \phi$:
consider a  contour integral $\oint \nabla \phi \cdot d\bl$ along the boundaries of the unit cell.
Since there are $N$ vortices inside the contour, the integral must
equal $2\pi N$. On the other hand, the assumption of $\nabla \phi$
 being periodic would have resulted in vanishing of the integral due to the cancellation between the
contributions of the opposite edges. 

Although the phase cannot be made periodic, it is quasi-periodic and can be
expressed in a closed form through Weierstrass sigma-function $\sigma(z)$.
Since (\ref{vdef}) yields
$$
\partial_y \phi+i\partial_x \phi =\sum_i \left(\zeta(z-Z_i)+\frac{\pi
\overline{Z_i}}{l^2}\right)~~,
$$
the difference
$\phi(\br)- \phi(\br_0)$, where $\br_0$ is an arbitrary reference point,  can be written as
\begin{multline}
\int_{\br_0}^{\br} \nabla\phi \cdot d\bl 
= {\rm Im} \int_{z_0}^{z} (\partial_y\phi+i\partial_x\phi) dz
\\
={\rm Im}\sum_i \left(\int_{z_0-Z_i}^{z-Z_i} \zeta(w) dw 
+\frac{\pi \overline{Z_i}(z-z_0)}{l^2} \right),
\end{multline}
where the contours in all integrals are assumed to be the same.
For different contours, the equality holds modulo $2\pi$.
Since the Weierstrass sigma-function  $\sigma(z)$ is defined according to 
$\sigma'(z)/\sigma(z) = \zeta(z)$, which implies
$$
\ln\frac{\sigma(z)}{\sigma(z_0)} \equiv\int_{z_0}^{z}\zeta(z) dz\pmod{2\pi}~~,
$$
the phase $\phi(\br)$ can be written as
\begin{multline}
\phi(\br)- \phi(\br_0) \equiv
 \sum_i {\rm Arg}\frac{\sigma(z-Z_i)}{\sigma(z_0-Z_i)}\\
+\frac{\pi}{l^2}{\rm Im}\left((z-z_0)\sum_i \overline{Z_i}\right)
\pmod{2\pi}~~.
\label{phase1}
\end{multline}
Since the phase $\phi(\br)$ is defined up to an arbitrary constant, even
for a fixed gauge of the vector potential $\bA$,  it is convenient to
choose the constant $\phi(\br_0)$ in such a way that
\begin{equation}
\phi(\br)= \sum_i {\rm Arg} [\sigma(z-Z_i)]
+\frac{\pi}{l^2}{\rm Im}\left(z\sum_i \overline{Z_i}\right)
\pmod{2\pi}~~.
\label{explicitphase}
\end{equation}
Eq-n (\ref{explicitphase}) is the explicit expression for the phase $\phi(\br)$ used in
the main text and in the rest of the appendix.
\subsection{Two vortices per unit cell}
In general, the procedure of separation of the phase $\phi(\br)$ into $\phi_A$ and
$\phi_B$ is straightforward by using (\ref{explicitphase}). In this section
we illustrate it for the simplest  case  of $N=2$ with two vortices  located 
at $\bR_A$ and $\bR_B$ inside an arbitrary reference unit cell of size
$l\times l$ (see Fig.\ref{fig:unitcell}). The  vector potential in the
symmetric gauge is given by
$$
\frac{e}{c}\bA = \frac{\pi}{l^2}(-y,x)~~,
$$
and the velocities $\bv_{A,B}$ can be defined similarly to $\bv$:
\begin{equation}
\bv_{A(B)}(\br) 
= 
\frac{2i \pi \hbar}{l^2}
\sum_{\bQ\neq 0} \frac{(Q_y,-Q_x)}{Q^2}
e^{i\bQ\br} e^{-i\bQ \bR_{A(B)}}~~.
\label{superv2AB}
\end{equation}
%
Clearly, $\bv_{A(B)}$ automatically satisfy  $\bv_A+\bv_B = 2\bv$ (see (\ref{vfour})).
Moreover, since $\bv_{A(B)}/2$ is formally given by the same
Fourier expansion as $\bv$ for $N=1$ (cf. (\ref{vfour})),  we
have
\begin{equation} 
\frac{\bv_{A(B)}}{2}=\frac12\nabla \phi_{A(B)} -\frac{e}{2c}\bA~~.
\end{equation}
where $\phi_{A(B)}$ satisfies the analogue of (\ref{deltacurl}):
\begin{equation}
\label{deltacurlA}
\nabla\times\nabla
\phi_{A(B)}=2\pi\hat{z}\sum_{\btau}{\delta(\br-\bR_{A(B)}-\btau)}~~,
\end{equation}
The closed form expressions for $\bv_{A(B)}$ can be read off directly from
the formulae of the previous section. After defining
\begin{align*}
F(\br) &= {\rm Arg}[\sigma(x+iy)]\\
\bZ(\br) &= \left({\rm Im} \zeta(z) +\frac{\pi y}{l^2}, {\rm Re} \zeta(z)-
\frac{\pi x}{l^2}\right)~~,
\end{align*}
where $\sigma(z)$ and $\zeta(z)$ are Weierstrass functions with periods
$(l,il)$, the superfluid velocities  are simply $\bv_{A(B)} =
\bZ(\br-\bR_{A(B)})$. The phases $\phi_{A(B)}$  are given by
\begin{equation}
\phi_{A(B)}(\br) = F(\br-\bR_{A(B)})+\frac{e}{c}\bA(\bR_{A(B)})\cdot \br~~;
\end{equation}
by construction they obey the condition $\phi_A(\br)+\phi_B(\br)=\phi(\br)$.

In our discussion of the quasiparticle spectrum,
we use the symmetry properties of the  phase difference $\dphi=\phi_A(\br)-\phi_B(\br)$.
Let us show that after a translation by the primitive lattice vector
$l\hat{x}$, the phase difference acquires a $\pi$-shift. Consider the following integral along a
straight line connecting $\br$ and $\br+l\hat{x}$:
$$
I(\br)= \int_{\br}^{\br+l\hat{x}}(\bv_A-\bv_B) \cdot d\bl~~.
$$
Functions $\bv_{A(B)}$ are periodic in unit cell and therefore
$I(x,y)$ does not depend on $x$:
$$
I(\br)=  \int_0^{l}[\bv_A(\xi,y)-\bv_B(\xi,y)] d\xi~~.
$$
Next, we note that the contour integral along the path shown in
Fig.\ref{fig:proof}  is a multiple of $2\pi$:
\begin{equation}
\oint (\bv_A-\bv_B)\cdot d\bl =
\oint (\nabla \phi_A-\nabla\phi_B)\cdot  d\bl = 2\pi (n_A-n_B)~,
\label{contint}
\end{equation}
where the  $n_A-n_B$ is the difference between the  number of $A$ and $B$ vortices inside the
contour. Since the integral along the vertical edges vanishes due to the 
periodicity of superfluid velocities $\bv_{A(B)}$, we find that 
$$
I(x,y_1)-I(x,y_2) = 2\pi n~~.
$$
Choosing $y_1$ and $y_2$ so that the horizontal segments are symmetrically
located around vortex $A$ as shown in Fig.\ref{fig:proof}, we have
$I(x,y_1)=-I(x,y_2)$ which yields $I(x,y_1)=\pi$,  and consequently,
\begin{equation}
\delta\phi(\br+l\hat{x})-\delta\phi(\br)
\equiv \pi \pmod{2\pi}~~.
\end{equation}
Combining this result with a similar identity for the translations in
$y$-direction:
$$
\delta\phi(\br+l\hat{y})-\delta\phi(\br)
\equiv \pi \pmod{2\pi}~~.
$$
we find that  $\exp(i\delta\phi(\br))$ can be written as  a product of
$\exp(i\pi(x+y)/l)$ and a periodic function with a unit cell $l\times l$.

Other useful identities involving  $\delta\phi=\phi_A-\phi_B$ are
\begin{align}
\dphi(x,y)+\dphi(l/2-x,y)&=\pi/2~~,\\
\dphi(x,y)+\dphi(x,l/2-y)&=-\pi/2~~,\\
\dphi(x,y)+\dphi(-x,-y)&=0~~.
\end{align}

\begin{table*}
\begin{minipage}{\columnwidth}
\begin{tabular}{|l|r|rr|l|}
\hline
$g$&$\Psi'=D(g)\Psi$ & $E'$ &$\bk'$ & $\theta_{1(2)}'$\\
\hline
\hline
$1$&
$\Psi(\br)$&
$E$,&
$\bk$&
$ \theta_{1(2)}$
\\
\hline
$m_x$&
$e^{-i\dphi} \sigma_1\Psi(g\br)$&
$E$,&$\bpi+(-k_x,k_y)$&
$ \pi/2+\theta_{1(2)}$
\\
\hline
$m_y$&
$e^{-i\dphi} \sigma_1\Psi(g\br)$&
$-E$,&$\bpi+(k_x,-k_y)$&
$ \pi-\theta_{1(2)}$
\\
\hline
$I$&
$\Psi(g\br)$&
$-E$,&$-\bk$&
$ \pi/2-\theta_{1(2)}$
\\
\hline
$P$&
$e^{-i\dphi} \Psi(g\br)$&
$-E$,&$\bpi-\bk$&
$ \pi/2-\theta_{2(1)}$\\
\hline
$P m_x$&
$\sigma_1 \Psi(g\br)$&
$-E$,&$(k_x,-k_y)$&
$ -\theta_{2(1)}$\\
\hline
$P m_y$&
$ \sigma_1 \Psi(g\br)$&
$E$,&$(-k_x,k_y)$&
$ \pi/2+\theta_{2(1)}$\\
\hline
$PI$&
$e^{-i\dphi} \Psi(g\br)$&
$E$,&$\bpi+\bk$&
$ \theta_{2(1)}$\\
\hline
\end{tabular}
\centering
\medskip

(a) 
\end{minipage}%
\begin{minipage}{\columnwidth}
\begin{tabular}{|l|r|rr|l|l|}
\hline
$Cg$&$D(Cg)\Psi=\Psi'$ & $E'$ &$\bk'$ & $\theta_{1(2)}'$\\
\hline
\hline
$C$&
$e^{-i\dphi}\sigma_2\Psibar(\br)$&
$E$,&$\bpi-\bk$&
$ \pi/2+\theta_{1(2)}$\\
\hline
$Cm_x$&
$\sigma_3\Psibar(g\br)$&
$E$,&$(k_x,-k_y)$&
$ \theta_{1(2)}$\\
\hline
$Cm_y$&
$\sigma_3\Psibar(g\br)$&
$-E$,&$(-k_x,k_y)$&
$ \pi/2-\theta_{1(2)}$\\
\hline
$CI$&
$e^{-i\dphi}\sigma_2\Psibar(g\br)$&
$-E$,&$\bpi+\bk$&
$ -\theta_{1(2)}$\\
\hline
$CP$ &
$\sigma_2 \Psibar(g\br)$&
$-E$,&$\bk$&
$ -\theta_{2(1)}$\\
\hline
$CP m_x$&
$e^{-i\dphi}\sigma_3 \Psibar(g\br)$&
$-E$,&$\bpi+(-k_x,k_y)$&
$ \pi/2-\theta_{2(1)}$\\
\hline
$CP m_y$&
$e^{-i\dphi}\sigma_3 \Psibar(g\br)$&
$E$,&$\bpi+(k_x,-k_y)$&
$ \theta_{2(1)}$\\
\hline
$CPI$&
$\sigma_2 \Psibar(g\br)$&
$E$,&$-\bk$&$
 \pi/2+\theta_{2(1)}$\\
\hline
\end{tabular}
\centering
\medskip

(b)
\end{minipage}
\caption{\label{sympsi}Transformation of the wavefunctions and the self-adjoint extensions
under various symmetry operations. The right column contains operations
involving complex conjugation.}
\end{table*}


\section{Two Aharonov-Bohm fluxes -- exact solutions}
\subsection{Elliptic coordinates and Mathieu functions}
In this appendix, we analyze the properties of a two-component Dirac particle
moving in the field of two Aharonov-Bohm fluxes. We will prove that
the exact solution of the problem is not unique and depends on 
boundary conditions imposed near the fluxes.
To achieve this goal, we first derive a  method allowing us to construct exact
solutions for the problem of a {\em Schr\"odinger} particle in the field of
two Aharonov-Bohm solenoids, which  carry arbitrary 
fractional fluxes $\alpha\phi _{0}$ and $\beta\phi_{0}$, 
with $\phi_0=hc/e$. We illustrate our procedure by analyzing
several simple,  physically interesting examples 
with  $\alpha,\beta=\pm\frac12$. 
After solving several model problems for a Schr\"odinger particle in the
field of two fluxes, we apply the technique 
to the case of two-component {\em Dirac}
particles. 

\begin{figure}
\includegraphics[width=0.4\textwidth]{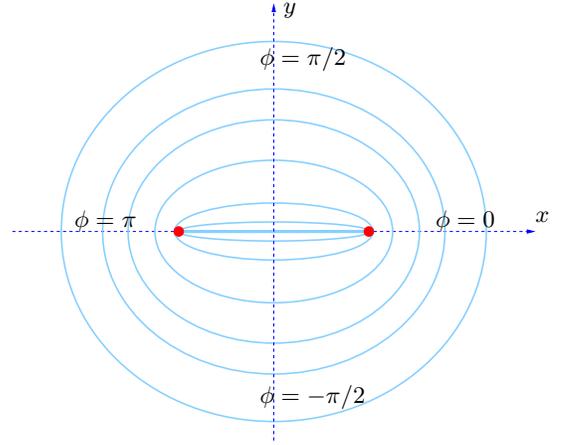}
\caption{\label{Fig:elliptic} Elliptic coordinate system. Each
line of constant $z$ is an  ellipse. At $z=0$ 
the ellipse is reduced to a segment $(-a,a)$ of
the real axis.}
\end{figure}

In what follows, we will extensively use Mathieu functions.  Therefore, we
start by reminding the reader of some
of their essential properties: the details can be found in Refs.
\onlinecite{mclachlan}, \onlinecite{morse}, and \onlinecite{stegun}.
Mathieu functions appear naturally as solutions 
of problems possessing elliptic
symmetry and expressed in elliptical coordinates. 
The most common examples are Laplace and Helmholtz equations with
boundary conditions specified on an ellipse.
The elliptic coordinates are defined according to
\begin{align}
x&=a\cosh z\cos \phi~~,\\
y&=a\sinh z\sin \phi~~,
\end{align}
where $a$ is a constant. Points with $z=0$  lie on a 
segment $(-a,a)$ traversed twice  when $\phi$ is
increased from $0$ to $2\pi$: first from $+a$ to $-a$, and 
then backwards. For
finite $z$ the contours $z=\rm const$ are ellipses (see
Fig.\ref{Fig:elliptic}):
$$
\frac{\displaystyle x^2}{a^2\cosh^2z}+\frac{y^2}{a^2\sinh^2z}=1~~,
$$
and angular coordinate $\phi$ is used to unambiguously specify the
position of the point on the ellipse of constant $z$.
Each ellipse is described by a major and minor semi-axes of 
lengths $a\cosh z$ and $a\sinh z$ respectively. 
The linear eccentricity defined as the distance from the center
to either focus, is the same for the entire family of ellipses and equals $a$. 
At large distances $r \gg a$, the contours of constant $z$ reduce to circles,
and the elliptic coordinates effectively reduce to the ordinary polar
coordinates $(\ln\rho,\theta)$.
\psfrag{f_x}{$x$}
\psfrag{f_y}{$y$}
\psfrag{f_0}{$\phi=0$}
\psfrag{f_1}{$\phi=\pi/2$}
\psfrag{f_2}{$\phi=\pi$}
\psfrag{f_3}{$\phi=-\pi/2$}
Both the Poisson and Helmholtz equations are separable in these
coordinates -- this is why the elliptic coordinate 
system is useful in various problems of
mathematical physics in the first place. The Laplacian in elliptic
coordinates assumes the following form:
$$
\frac{\partial ^{2}}{\partial x^{2}}+\frac{\partial ^{2}}{\partial y^{2}}=
\frac{\displaystyle 2}{a^{2}(\cosh 2z-\cos 2\phi )}\left(\frac{\partial
^{2}}{\partial z^{2}}+\frac{\partial ^{2}}{\partial \phi ^{2}}\right)~~,
$$
and the Helmholtz equation $\nabla^2 f+k^{2}f=0$ becomes
$$ 
\left(\frac{\partial ^{2}}{\partial z^{2}}+\frac{\partial ^{2}}{\partial
\phi ^{2}}\right)f+\frac{a^{2}k^{2}}{2}(\cosh 2z-\cos 2\phi )f=0~~.
$$
After  separation of variables via $f=F(z)G(\phi)$, one obtains
$$
\frac{1}{F(z)}\frac{d^{2}F}{dz^{2}}+\frac{1}{G(\phi )}\frac{d^{2}}{d\phi ^{2}}+
\frac{a^2k^2}{2}(\cosh 2z-\cos 2\phi )=0~~,
$$
which results in
\begin{align}
\frac{1}{F(z)}\frac{d^{2}F}{dz^{2}}&+\frac{a^{2}k^{2}}{2}\cosh 2z=A~~, \\
\frac{1}{G(\phi )}\frac{d^{2}}{d\phi ^{2}}&-\frac{a^{2}k^{2}}{2}\cos
2\phi=-A~~,
\end{align}
where $A$ is the separation constant. The conventional form of these 
equations, 
which of course must be solved simultaneously for the same constant $A$, is 
\begin{align}
\frac{d^{2}G}{d\phi ^{2}}&+(A-2q\cos 2\phi )G(\phi )=0\label{separatedG}~~,\\
\frac{d^{2}F}{dz^{2}}&-(A-2q\cosh 2z)F(z)=0~~,\label{separatedF}
\end{align}
where $q$ is a parameter defined as $q=a^2k^2/4$.
\subsubsection{Mathieu functions}
Since the solutions must be single-valued functions, $G(\phi)$ must be
$2\pi$-periodic. This condition restricts $A$ to a discrete set of
so-called characteristic values, which is traditionally written as a union
of two subsets representing solutions that are even or odd under
transformation $\phi\to-\phi$:
$$
\left\{ 
\begin{array}{rrrrrr}
a_{0},&a_{1},&a_{2},&a_{3},&...&{\text{(even solutions)}}\\
& b_{1},&b_{2},&b_{3},&...&{\text{ (odd solutions)}}
\end{array}
\right.~~.
$$
While working with the elliptic coordinates, it is useful to keep in mind
similarities and distinctions from the polar coordinate system. In the
latter, the equation for the angular component assumes  a form $G''+A G =0$,
and the requirement of $2\pi$-periodicity restricts  $A$ to a set of  $m^2$
with integer $m$, irrespectively of $k$. In the elliptic case, this  no
longer holds, and the characteristic values 
$a_j$, $b_j$ depend on $q=a^2k^2/4$.

For $A=a_{n}$ the periodic solution of the equation for $G(\phi)$ is an
even function of $\phi$, which is denoted as $ce_{n}(\phi ,q)$, $n=0,1,2,...$.
The odd solutions, occurring for  $A=b_{n}$,  are denoted as $se_{n}(\phi ,q)$.
Functions $ce_n$ and $se_n$ are known as Mathieu functions, the notation is
to remind us that they reduce to $\cos(n\phi)$ and $\sin(n\phi)$ respectively 
in the limit $q\to 0$.

The equation for $G$ is of second order, so for a given $A=a_{n}$ (or $A=b_{n}$)
there is a second independent solution, in addition to $ce_n(\phi,q)$ (or
$se_n(\phi,q)$). However, the second solution of the angular equation 
is {\em never} periodic, and therefore it
is of limited, if any, importance in  applications.
The  characteristic values $a_{n}$, $b_{n}$,  as well as the Mathieu
functions $ce_{n}$ and $se_{n}$  are tabulated and well studied, see for
example Ref. \onlinecite{mclachlan}.  Finally, we note that Mathieu functions
$\{ce_n(\phi,q),se_n(\phi,q)\}$ form a complete set in the 
interval $(0,2\pi)$ and satisfy the following
orthonormality relations:
\begin{align}
\int_{0}^{2\pi}ce_{n}(\phi,q)ce_{m}(\phi,q)\, d\phi &=\delta_{nm}~~,\\
\int_{0}^{2\pi}se_{n}(\phi,q)se_{m}(\phi,q)\, d\phi &=\delta_{nm}~~,\\
\int_{0}^{2\pi}ce_{n}(\phi,q)se_{m}(\phi,q)\, d\phi &=0~~.
\end{align}


\begin{table}[htb]
{\centering
\begin{tabular}{c|c|c|c}
\hline
\hline
A& 1st solution&$f(0,q)$&$ \partial_z f(0,q)$\\
\hline
$a_{n}$&$Ce_{n}(z,q)\propto Mc^{(1)}(z,q)\propto Je_{n}(z,q)$&$\neq0$&$ 0 $\\
$b_{n}$&$Se_{n}(z,q)\propto Ms^{(1)}(z,q)\propto Jo_{n}(q,z)$&$0$    &$\neq0$\\
\hline
A&  2nd solution&$f(0,q)$&$\partial_z f(0,q)$\\
\hline
$a_{n}$&$Fey_{n}(z,q)\propto Mc^{(2)}(z,q)\propto Ne_{n}(z,q) $ & $\neq 0$&$\neq 0 $\\
$b_{n}$&$Gey_{n}(z,q)\propto Ms^{(2)}(z,q)\propto No_{n}(z,q) $ & $\neq
0$&$\neq 0$\\
\hline
\hline
\end{tabular}}
\caption{\label{Table:Notation}Modified Mathieu functions: notation and values at $z=0$.}
\end{table}



\subsubsection{Modified Mathieu functions}
Now we turn to the equation (\ref{separatedF}) describing  the ``radial''
part of the solutions. Since (\ref{separatedF}) and (\ref{separatedG}) must
be solved simultaneously for the same value of parameter $A$, we are
interested in solutions of (\ref{separatedF}) only for $A=a_{n}$ or $b_{n}$.
These solutions are known as Modified Mathieu functions. For every value of
parameter $A$ there are two linearly independent solutions: for $A=a_n$
these solutions are denoted as $Je_n(z,q)$ and $Ne_n(z,q)$, and for $A=b_n$
the solutions are  $Jo_n(z,q)$ and $No_n(z,q)$. These functions 
take place of Bessel $J_m(r)$ and Neumann $N_m(r)$ 
functions in the more familiar
case of polar coordinates; the  letters ``e'' and ``o'' stand for ``odd''
and ``even'' respectively.

Unfortunately  the  notation used  for Modified Mathieu functions is not
standardized, and for convenience we provide the Table \ref{Table:Notation}, which
relates notations used by different authors. The functions of the first
kind, $Je_n(z,q)$, $Jo(z,q)$,  are proportional to  the Mathieu functions of
imaginary argument $ce_n(iz,q)$ and $se(iz,q)$. In our subsequent analysis,
we will also use the properties of the Modified Mathieu functions at $z=0$
and $z\gg 1$,
which  are summarized  in the right two columns of Table
\ref{Table:Notation} and Table \ref{Table:asympt} respectively.
\begin{table}
\begin{tabular}{llrl}
\hline
\hline
$ Je_{2n}(z,q)$  &$\approx$&$  p_{2n}  \gamma   e^{-z/2}$  &$\sin (\sqrt{q}e^{z}+\frac{\pi }{4}) $\\
$ Ne_{2n}(z,q)$  &$\approx$&$ -p_{2n}   \gamma  e^{-z/2}$  &$\cos (\sqrt{q}e^{z}+\frac{\pi }{4})  $\\
$ Je_{2n+1}(z,q)$&$\approx$&$ -p_{2n+1} \gamma e^{-z/2}$  &$\cos(\sqrt{q}e^{z}+\frac{\pi }{4}) $\\
$ Ne_{2n+1}(z,q)$&$\approx$&$ -p_{2n+1} \gamma e^{-z/2}$  &$\sin(\sqrt{q}e^{z}+\frac{\pi }{4}) $\\
\hline
$ Jo_{2n}(z,q)$  &$\approx$&$  s_{2n}  \gamma  e^{-z/2}$  &$\sin (\sqrt{q}e^{z}+\frac{\pi}{4}) $\\
$ No_{2n}(z,q)$  &$\approx$&$ -s_{2n}  \gamma  e^{-z/2}$  &$\cos (\sqrt{q}e^{z}+\frac{\pi }{4}) $\\
$ Jo_{2n+1}(z,q)$&$\approx$&$ -s_{2n+1}\gamma e^{-z/2}$  &$\cos(\sqrt{q}e^{z}+\frac{\pi }{4}) $\\
$ No_{2n+1}(z,q)$&$\approx$&$ -s_{2n+1}\gamma e^{-z/2}$  &$\sin(\sqrt{q}e^{z}+\frac{\pi }{4})$\\
\hline
\hline
\end{tabular}
\caption{\label{Table:asympt}Asymptotic expressions for Modified Mathieu
functions for $z\gg 1$ in terms of $\gamma=2^{1/2}(\pi^2 q)^{-1/4}$ and the standard coefficients $p_n$ and $s_n$
(see Ref. \onlinecite{mclachlan}).}
\end{table}
In the most general case, the solution of the Helmholtz equation can be
written in  the following form:
\begin{multline}
f=\sum _{n=0}ce_{n}(\phi ,q) \Bigl(\alpha _{n}Je_{n}(z,q)+\beta
_{n}Ne_{n}(z,q)\Bigr)\\
+\sum _{n=1}se_{n}(\phi ,q) \Bigl(\gamma _{n}Jo_{n}(z,q)+\delta
_{n}No_{n}(z,q)\Bigr)~~,
\label{gencase}
\end{multline}
where constants \( \alpha _{n},\beta _{n},\gamma _{n},\delta _{n} \) depend 
on the  boundary conditions. In addition to 
the external boundary conditions, additional 
attention should be paid to  the segment $(-a,a)$
corresponding to $z=0$: ordinarily the solutions and their derivatives are
continuous across the segment providing a restriction on the coefficients
$\alpha_n{\ldots} \delta_n$. We, however, will be also interested in
solutions that experience discontinuity across the line, which will provide
a different set of constraints on the coefficients in (\ref{gencase}).
\subsubsection{Model Schr\"odinger problem: a particle inside an elliptic box}
We begin with the wave equation describing a particle inside an impenetrable  box of
elliptical shape. The wavefunction $\psi$ is assumed to be 
continuous everywhere inside the ellipse, together with its first derivatives.

It is easy to see that the four products  in (\ref{gencase}) and their
normal derivatives  have the following
continuity properties across the segment $(-a,a)$:

\vskip .1truein
\par
\begin{tabular}{l|c|c}
\hline
\hline
$F$& $F$ continuous? & $\partial_n F$ continuous?\\
\hline
\( ce_{n}(\phi ,q) Je_{n}(z,q) \)& Yes&Yes\\
\( se_{n}(\phi ,q) Jo_{n}(z,q) \)& Yes& Yes\\
\( ce_{n}(\phi ,q) Ne_{n}(z,q) \)& Yes& No\\
\( se_{n}(\phi ,q) No_{n}(z,q) \)& No& Yes\\
\hline
\hline
\end{tabular}
\par
\vskip .1truein

For a particle-in-the-box problem, therefore, the solution must be of the form
$$
f=\sum _{n=0}\alpha _{n} ce_{n}(\phi ,q) Je_{n}(z,q)+
\sum _{n=1}\gamma _{n} se_{n}(\phi ,q) Jo_{n}(z,q).
$$
Coefficients \( \alpha _{n} \) and \( \gamma _{n} \) are determined by the
boundary conditions imposed on the exterior boundary of the system.
If the boundary of the ellipse is described by $z=R$, then the eigenstates
can be labelled by two integer indices: the angular index  $n$, the radial
index $j$, and their parity under $\phi\to -\phi$. More explicitly, the
eigenfunctions are (up to normalization factors)
\begin{align}
\psi^{(e)}_{nj}(\phi,z) &= ce_n(\phi,q^{e}_{nj}) Je_n(\phi,q^{e}_{nj})~~,\\
\psi^{(o)}_{nj}(\phi,z) &= se_n(\phi,q^{o}_{nj}) Jo_n(\phi,q^{o}_{nj})~~,
\end{align}
where $q^{e}_{nj}$ (or $q^{o}_{nj}$) is defined as the $j$-th solution of
$Je_n(R,q)=0$ (or $Jo_n(R,q)=0$).
The corresponding  energy eigenvalues are 
\begin{equation}
E_{nj}^{o(e)}=\frac{\displaystyle \hbar ^{2}}{\displaystyle 2m}k^{2}=
\frac{\displaystyle\hbar ^{2}}{\displaystyle
2m}\frac{\displaystyle4q^{o(e)}_{nj}}{\displaystyle a^{2}}~~.
\end{equation}

\subsection{Schr\"odinger particle in the field of two Aharonov-Bohm half-fluxes}
\subsubsection{Boundary condition on a segment $z=0$}
To solve the Schr\"{o}dinger equation
\begin{equation} 
\left(-i\nabla -eA\right)^{2}\psi +k^{2}\psi =0
\end{equation}
with the vector potential \( A({\br}) \) describing two Aharonov-Bohm fluxes
at positions \( (a,0) \) and \( (-a,0) \), we perform singular gauge transformation 
\begin{equation}
\psi \rightarrow f \exp \left(ie\int
_{\mathbf{r}_{0}}^{\mathbf{r}}A(\br) \cdot d\br\right)~~.
\end{equation}
Since the integral in the exponent depends on the path of integration, we need to
introduce a branch cut in order to obtain single-valued functions. We chose
the segment  $(-a,a)$ of the real axis as the branch cut:
\begin{equation}
\label{transform}
\psi \rightarrow f \exp \left(\frac{i\theta_{1}+i\theta _{2}}{2}\right)~~,
\end{equation}
where $-\pi \le \theta _{1}<\pi$ and $0\le \theta _{2}<2\pi$. This
transformation eliminates the vector potential, and the resulting
equation is just the  Helmholtz equation
$$
\nabla ^{2}f+k^{2}f=0~~,
$$
except that the solutions must have a branch cut on a segment  $(-a,a)$:
$$
f(x, y+\epsilon) = -f(x, y-\epsilon )\textrm{ for }x\in (-a,a)~.
$$
Elliptic coordinates provide a natural framework for imposing the boundary
condition:
$$
f(z=0,\phi )=-f(z=0,-\phi)~~,
$$
i.e. at $z=0$ the solution $f(z=0,\phi)$ must be an odd function
of $\phi$.

Starting from a general form
\begin{multline}
f=\sum _{n=0}ce_{n}(\phi ) [\alpha _{n}Je_{n}(z)+
\beta _{n}Ne_{n}(z)]\\
+\sum _{n=1}se_{n}(\phi ) [\gamma _{n}Jo_{n}(z)+
\delta _{n}No_{n}(z)]~~,
\end{multline}
we obtain for $f(z=0, \phi)$:
\begin{equation*}
\sum _{n=0}ce_{n}(\phi ) [\alpha _{n}Je_{n}(0)
+
\beta _{n}Ne_{n}(0)]+\sum _{n=1}\delta _{n} se_{n}(\phi ) No_{n}(0),
\end{equation*}
which implies the following relation between \( \alpha _{n} \) and
\( \beta _{n} \):
\begin{equation}
\label{ab}
\alpha _{n}Je_{n}(0)+\beta _{n}Ne_{n}(0)=0~~.
\end{equation}
Similarly, the requirement for \( \psi (x,y) \) to have continuous
normal derivative across the branch cut yields
$$
\frac{\partial }{\partial z}f(z,\phi )\left|_{z=0}\right. 
=\frac{\partial }{\partial z}f(z,-\phi )\left| _{z=0}\right.
~~.
$$
Since $\frac{\partial }{\partial z}f(z,\phi )$ at $z=0$ equals
\begin{multline}
\sum _{n=0}ce_{n}(\phi ) [\alpha _{n}Je_{n}'(0)+\beta
_{n}Ne_{n}'(0)]\\+\sum _{n=1}se_{n}(\phi ) [\gamma _{n}Jo_{n}'(0)+\delta
_{n}No_{n}'(0)]~~,
\end{multline}
we obtain
\begin{equation}
\label{gd}
\gamma _{n}Jo_{n}'(0)+\delta _{n}No_{n}'(0)=0~~.
\end{equation}
Equations (\ref{ab}) and (\ref{gd}) express the continuity of the
wavefunction $\psi(x,y)$ and its normal derivative at the segment $(-a,a)$.
As a result, the general expression for the wave function has the form
\begin{multline}
f(z,\phi)=
\sum _{n=0}\alpha _{n}(q) ce_{n}(\phi ,q)Le_n(z,q)\\
+\sum _{n=1}\gamma _{n}(q) se_{n}(\phi ,q)Lo_n(z,q)~~.
\label{gensolutionfluxes}
\end{multline}
To compactify the  notation,  we have introduced the following functions:
\begin{align}
Le_n(z,q)=\frac{Je_{n}(z,q)}{Je_{n} (0,q)}-\frac{Ne_{n}(z,q)}{Ne_{n} (0,q)}~~,\\
Lo_n(z,q)=\frac{Jo_{n}(z,q)}{Jo_{n}'(0,q)}-\frac{No_{n}(z,q)}{No_{n}'(0,q)}~~.
\end{align}
At this point one can pose and solve 
several problems with various external boundary
conditions, which we will consider next.
\psfrag{f_a}{$a$}
\psfrag{f_ma}{$-a$}
\psfrag{f_t1}{$\theta_1$}
\psfrag{f_t2}{$\theta_2$}
\begin{figure}
\includegraphics[width=0.35\textwidth]{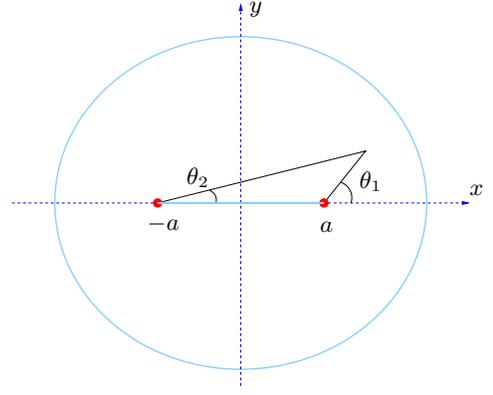}
\caption{\label{Fig:intproblem} Two half-fluxes inside an elliptical box.}
\end{figure}
\subsubsection{Schr\"odinger  particle in an elliptical box in the presence of two half-fluxes.}
The wavefunctions are zero on the boundary of the elliptical box \( z=R \),
and therefore
\begin{multline}
\sum _{n=0}\alpha _{n}(q) ce_{n}(\phi ,q)Le_n(R,q)\\
+\sum _{n=1}\gamma_{n}(q) se_{n}(\phi ,q)Lo_n(R,q)
=0.
\end{multline}
Using orthogonality of \( ce_{n}(\phi ) \) and \( se_{n}(\phi ) \), we find
that 
the eigenstates of are either even (labeled as ``$e$'') or odd (labeled as
``$o$'') functions of $\phi$, and their explicit form, up to
normalization factors, is given by
\begin{align}
f_{nj}^{(e)}(z,\phi)&=ce_{n}(\phi,q^{e}_{nj})Le_n(z,q^{e}_{nj})
~~,
\\
f_{nj}^{(o)}(z,\phi)&= se_{n}(\phi ,q^{o}_{nj})Lo_n(z,q^{e}_{nj})
~~,
\end{align}
where \( q^{e}_{nj} \) and \( q^{o}_{nj} \) are \( j \)-th solutions
of 
\begin{align}
\frac{Je_{n}(R,q)}{Je_{n}(0,q)}-\frac{Ne_{n}(R,q)}{Ne_{n}(0,q)}&=0~~,
\label{defeqn1}\\
\frac{Jo_{n}(R,q)}{Jo_{n}'(0,q)}-\frac{No_{n}(R,q)}{No_{n}'(0,q)}&=0
\label{defeqn2}
\end{align}
respectively. Therefore,  just as in the case of the no-flux problem of the
previous section, the eigenstates are classified according to parity
under $\phi\to-\phi$, and two
integers $(j,n)$, where $j=1,2,3,\ldots$, and $n=(0),1,2,\ldots$ for even (odd)
solutions. The energy  of the eigenstates $|n,j,o(e)\rangle$ is
\begin{equation}
E_{nj}^{o(e)}=\frac{\displaystyle \hbar ^{2}}{\displaystyle 2m}k^{2}=
\frac{\displaystyle\hbar ^{2}}{\displaystyle
2m}\frac{\displaystyle4q^{o(e)}_{nj}}{\displaystyle a^{2}}~~.
\end{equation}
\begin{figure}
\includegraphics[width=0.35\textwidth]{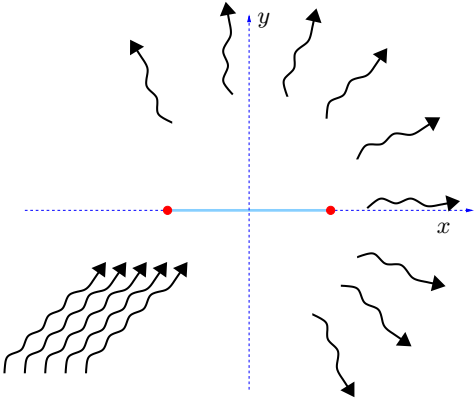}
\caption{Two half-fluxes: scattering problem.}
\end{figure}
\subsubsection{Scattering of a Schr\"odinger particle by two fluxes}
Now we consider the problem of a particle scattered by two fluxes $\phi_0/2$
and $-\phi_0/2$. To find the scattering cross-section \( \sigma (\hat{\mathbf{n}},\, \hat{\mathbf{n}}') \)
and the scattering amplitude \( F(\hat{\mathbf{n}},\, \hat{\mathbf{n}}') \) we
need to match general solution $f(z,\phi; q)$ given 
by (\ref{gensolutionfluxes})
with the boundary conditions at infinity that contains only the incoming and
the scattered waves:
\begin{equation}
f_{\hat{\mathbf{n}}}=e^{i\mathbf{k}\cdot
\mathbf{r}}+\frac{\exp(ikr)}{\sqrt{r}}F(\hat{\mathbf{n}},\,
\hat{\mathbf{n}}')~~.
\label{scatt}
\end{equation}
Here, \( \hat{\mathbf{n}}=(\cos \theta , \sin \theta) \) specifies the direction
of the incident wave and \( \hat{\mathbf{n}}'=(\cos \phi, \sin \phi) \)
describes the direction of scattered wave. 
The first condition, Eq.  (\ref{gensolutionfluxes}), encodes the information about
the presence of two half-fluxes, while the second 
condition, Eq. (\ref{scatt}), restricts the
the wavefunctions to the sum of the incident 
plane wave $\exp(i\bk\cdot \br)$
and the outgoing wave described by the 
second term. Together, they allow
to determine the scattering amplitude 
unambiguously. To match the two
expressions, we write (\ref{scatt}) 
in terms of Mathieu functions. Expansion of a plane
wave $e^{i\mathbf{k}\cdot \mathbf{r}}$ in 
Mathieu functions basis has the following form:
\begin{multline}
\label{planewave}
e^{ik(x\cos \theta +y\sin \theta )}=
\sum _{n=0}\rho _{n}(q)Ce_{n}(z)ce_{n}(\phi )ce_{n}(\theta )\\
+\sum _{n=1}\sigma _{n}(q)Se_{n}(z)se_{n}(\phi )se_{n}(\theta )~~,
\end{multline}
where the 
coefficients \( \rho _{n}(q),\sigma _{n}(q) \) are 
related to the known standard
factors \( p_{n}(q),s_{n}(q) \) from the theory of Mathieu functions  (cf.
Table \ref{Table:asympt}) as
\begin{align}
\rho _{2n}=\frac{1}{p_{2n}}~,  &  \quad \rho_{2n+1}=\frac{i}{p_{2n+1}}~,\\
\sigma _{2n}=\frac{1}{s_{2n}}~, & \quad \sigma_{2n+1}=\frac{i}{s_{2n+1}}~.
\end{align}
Now we turn to the scattered wave: the  most general expression for the
solution, which contains only the outgoing wave, is given by
\begin{multline}
\label{scattered}
\frac{e^{ikr}}{\sqrt{r}}F(\hat{\mathbf{n}},\, \hat{\mathbf{n}}') =\sum _{n=0}\lambda _{n}ce_{n}(\phi )ce_{n}(\theta )He^{(1)}_{n}(z)\\+
\sum _{n=1}\mu _{n}se_{n}(\phi )se_{n}(\theta )Ho^{(1)}_{n}(z)~~.
\end{multline}
Functions \( He_{n}^{1}(z) \) and \( Ho_{n}^{1}(z) \),  which play the role
of the  Hankel functions in the theory of Mathieu equation,  are defined as
\begin{equation}
\label{Hankel}
\left\{ \begin{array}{c}
He^{1}_{n}(z)=Je_{n}(z)+iNe_{n}(z)~~,\\
Ho^{1}_{n}(z)=Jo_{n}(z)+iNo_{n}(z)~~.
\end{array}\right.
\end{equation}
Combining (\ref{gensolutionfluxes}), (\ref{planewave}) and \ref{scattered} we find
\begin{align}
\lambda _{n}&=-\frac{\displaystyle Je_{n}(0,q)}{\displaystyle
Je_{n}(0,q)+iNe_{n}(0,q)}\rho _{n}~~,\\
\mu _{n}&=-\frac{\displaystyle Jo_{n}'(0,q)}{\displaystyle
Jo_{n}'(0,q)+iNo_{n}'(0,q)}\sigma _{n}~~.
\label{mulambda}
\end{align}
We remind the reader that all quantities on the right hand side of
(\ref{mulambda}) are known, and to obtain the scattering amplitude
$F(\hat{\mathbf n}, \hat{\mathbf n'})$ from (\ref{scattered}) we only need to
use the long distance expressions for $He_n$ and $Ho_n$ that
are easy to find from Table \ref{Table:asympt}:
$$
\rho_n He_{n}  \approx \sigma_n Ho_n \approx   \left(\frac{4}{\pi^2{q}}\right)^{1/4}
e^{-z/2+i\sqrt{q}e^{z}-i\pi /4}~~.
$$
At large distances, we have  $e^{z}=2r/a$, the elliptic 
coordinate $\phi$ is reduced to the  ordinary
polar angle, and therefore
\begin{equation}
\rho_{n} He_{n}\approx \sigma_{n}Ho_{n} \approx \left(\frac{a^2}{\pi^2 q}\right)^{1/4}e^{-i\pi /4}\frac{1}{\sqrt{r}}e^{ikr}~.
\end{equation}
Finally, substitution of these expressions into (\ref{scattered}) yields the
following exact expression for the scattering amplitude:
\begin{multline}
F(\theta ,\phi )=-\left(\frac{a^2}{\pi^2 q}\right)^{1/4}e^{-i\pi /4}
\times \\
\left( \sum _{n=0}\frac{Je_n(0,q)}{He_n(0,q)}ce_{n}(\phi
,q)ce_{n}(\theta ,q)
\right.
\\
\left.
+
\sum _{n=1}\frac{Jo_n'(0,q)}{Ho'_n(0,q)}se_{n}(\phi ,q)se_{n}(\theta ,q)\right)~.
\end{multline}

\subsection{Dirac equation}
\subsubsection{Relation between Schr\"odinger and Dirac problems}
Now we apply the technique we developed in this appendix to the 
problem of a Dirac particle moving in the
field of two half-fluxes, without specifying the external boundary condition
at this point. The Hamiltonian  $H_0$ describing such a system is
$$
\begin{pmatrix}
0 & (p_x-eA_x) -i (p_y-eA_y)\\
(p_x-eA_x) +i (p_y-eA_y) &0
\end{pmatrix},
$$
where $(p_x,p_y)$ is the momentum operator and ${\mathbf A}$ is the vector
potential of the two half-fluxes. Just as for the Schr\"odinger problem,
we perform a unitary transformation (\ref{transform}), which eliminates the
vector potential from the Hamiltonian, but at the expense
of introducing a branch cut. After choosing a 
segment of the real axis  $(-a,a)$ as the location of the branch
cut, the eigenfunctions of the Hamiltonian
$$
H=
e\begin{pmatrix}
0 & p_x -i p_y\\
p_x +i p_y &0
\end{pmatrix}
$$
can be found from the eigenfunctions of the Schr\"odinger problem, which we
described in the previous section: note that if $f=(u,v)^T$ is an
eigenfunction of $H$, then it is also an eigenfunction of 
$$
H^2 = 
\begin{pmatrix}
p_x^2 + p_y^2 &0\\
0 & p_x^2 + p_y^2
\end{pmatrix}~~.
$$
Thus, both $u$ and $v$  are solutions of the wave equation
\begin{equation}
(\nabla^2+E^2)u =0~~,
\label{schU}
\end{equation}
and  any solutions of our eigenvalue problem
\begin{align}
(p_x -i p_y)v &= Eu~~,\\
(p_x +i p_y)u &= Ev
\end{align}
can be written as
\begin{equation}
f=
\begin{pmatrix}
u(x,y)\\
\frac{1}{E} (p_x+ip_y)u(x,y)
\end{pmatrix}~~,
\label{constructEF}
\end{equation}
where $u(\br)$ is a solution of (\ref{schU}). The opposite statement
is also clearly valid: for any $u$, which satisfies (\ref{schU}),
wavefunction (\ref{constructEF}) is an eigenfunction of Hamiltonian $H$ with
energy $E$.  Thus, it naively appears that all solutions of the Dirac equation can be written as
(\ref{constructEF}). We will return to the validity of this statement in a moment, but
let us first apply (\ref{constructEF}) to a simpler problem of a {\em
single-flux} problem.
\subsubsection{The single-flux problem}
A problem of {\em Schr\"odinger} particle moving in the field of a single
half-flux located at the origin is most conveniently solved by eliminating the vector potential by means of
$\psi(\br)=\exp(i\theta/2)f(\br)$ transformation, where $\theta$ is the polar angle. The
resulting eigenvalue problem 
\begin{equation}
(\nabla^2+E^2)f =0
\label{auxschr}
\end{equation}
must be solved in a space of wavefunctions $f(r,\theta)$ that have a
branch-cut extending from $0$ to infinity. The cut can be chosen
arbitrarily, e.g., as a straight line $(0,+\infty)$, or $(-\infty,0)$.
The solutions are easily obtained in polar coordinates as
\begin{equation}
f = \sum_{-\infty}^{+\infty} e^{i(m+1/2)\theta}\Bigl[c_m J_{m+1/2}(kr)+ d_m
J_{-m-1/2}(kr) \Bigr].
\label{genJ}
\end{equation}
Consider now a class of problems where the boundary conditions allow the
particles to reach the origin --
this excludes problems where impenetrable walls of finite radius surround
the flux. 

The requirement of square-integrability of the wavefunctions requires that 
all coefficients $d_m,m\ge 0$ and $c_m, m< 0$ must be set to  zero except, possibly, $d_0$ and
$c_{-1}$. The eigenfunctions corresponding to  $c_{-1}$ and $d_0$ are
divergent at the origin, but square integrable.
Ordinarily, physical realization of the flux requires the
wavefunctions to be not only square-integrable, but also finite at $r=0$:
this eliminates the remaining arbitrariness and results in $d_m=0$ for all
$m\ge 0$, and $c_m=0$ for all $m<0$. Thus, for every angular channel $m$ there
is only one radial solution:
$$
\psi=e^{i\theta/2}f = \sum_{-\infty}^{+\infty} a_m e^{i m
\theta}J_{|m-1/2|}(Er)~~,
$$
 however, (\ref{auxschr}) is used as an auxiliary tool to obtain
solutions of  the Dirac problem via Eq. (\ref{constructEF}), requirement of
the wavefunctions being finite at $r=0$ is too restrictive: even if the upper
components $u$ of solutions are chosen to be  finite, then the lower
component of at least some wavefunctions will necessarily
be divergent, but square integrable at the origin. To consider the most
general situation, however, we choose $u$ is the form of (\ref{genJ})
allowing at this stage all square integrable wavefunctions. Using
(\ref{constructEF}) and the following expressions:
\begin{equation}
\partial_x\pm i\partial_y = e^{\pm i\theta}
\left(\partial_r\pm\frac{i}{r}\partial_{\theta}\right)~~,
\end{equation}
the solutions of the Dirac equation can  be written as
$$
f=\sum_m e^{i(m+\frac12)\theta} (c_m\chi_{m}^{(+)}+d_m \chi_{m}^{(-)}),
$$
where
\begin{equation}
\chi_{m}^{(\pm)} =
\begin{pmatrix}
J_{\pm(m+\frac12)}(Er)\\
-i e^{i\theta}
\left[
J_{\pm(m+\frac12)}'(Er) -\frac{(m+\frac12)}{Er}J_{\pm(m+\frac12)}(Er)
\right]
\end{pmatrix}~~.
\end{equation}
Using the standard identities for Bessel functions, $f$ can be written as
\begin{multline}
f=\sum_m e^{i(m+\frac12)\theta}
\left[c_m
\begin{pmatrix}
J_{m+\frac12}(Er)\\
i e^{i\theta}
J_{m+\frac32}(Er)
\end{pmatrix}\right.\\
\left.
+d_m 
\begin{pmatrix}
J_{-m-\frac12}(Er)\\
-i e^{i\theta}
J_{-m-\frac32}(Er) 
\end{pmatrix}
\right]
\end{multline}
For all values of $m\ge0 $ ($m\le -2$) the requirement of square-integrability
demands that $d_m=0$ ($c_m=0$). At $m=-1$, however there is an ambiguity. It
is impossible {\em a priori} to decide which of the two radial functions
\begin{equation}
\begin{pmatrix}
J_{-\frac12}(Er)\\
i e^{i\theta}
J_{\frac12}(Er)
\end{pmatrix}
\text{ or }
\begin{pmatrix}
J_{\frac12}(Er)\\
-i e^{i\theta}
J_{-\frac12}(Er) 
\end{pmatrix}
\end{equation}
 should be used. Either the upper or the lower component of the
wavefunction is divergent, but still square integrable. 
The attempt to set both
$c_{-1}$ and $d_{-1}$ to zero leads to the loss of completeness in the
angular basis: the wavefunctions described by $f\propto \exp(-i\theta/2)$  or,
equivalently, the original gauge wavefunctions $\psi$ 
which do not have angular dependence, would be
left out of the Hilbert space. On the other hand, to require that both
solutions are present in the spectrum independently would be too much: the
resulting set of basis functions is then overcomplete. 
As was shown by Jackiw and Gerbert, the solution is to use one linear
combination of the two solutions. Different regularizations of the problem
then correspond to different choice of the linear combination. Importantly,
not all
linear combinations are mathematically allowed, and can emerge from
the regularized problems: the relative phase of 
$c_{-1}$ and $d_{-1}$ turns out to be
fixed, and different boundary conditions, forming different self-adjoint
extensions of the problem are described through a single parameter $\theta$.
For a given $\theta$ the divergent part of the wavefunctions is
\begin{equation}
\frac1{\sqrt{r}}
\begin{pmatrix}
\sin \theta\\
-i e^{i\phi} \cos\theta
\end{pmatrix}~~.
\label{divergent1flux}
\end{equation}
Note that in this problem the basis functions have a useful property: both
the upper component and the lower component can be simultaneously written as 
separable functions, i.e. a product of two functions that depend only on
$\theta$ and only on $r$. This is a peculiarity of the single-flux problem,
and in general this property does not hold. Nevertheless, although the
analysis will be slightly more complicated, all essential properties of the
single-flux problem such as the necessity of additional boundary conditions
at flux locations and the form of the boundary conditions remain valid in
the case of two-flux problem as well.

\subsubsection{Dirac particle in the presence of two half-fluxes}
Once again, we use (\ref{constructEF}) to construct solutions of Dirac
equation from the solutions (\ref{gensolutionfluxes}) of the wave equation
in the presence of two half-fluxes.
In elliptical coordinates, 
$$
\partial_x+i \partial_y = \frac1{a\sinh z\cos \phi - a\cosh z\sin\phi}
\left(\partial_z +i\partial_{\phi}\right)~~.
$$
The  eigenfunctions of the Dirac equation obtained from
(\ref{constructEF}) and (\ref{gensolutionfluxes}) can be written now as
\begin{equation}
\sum _{n=0}\alpha _{n}(q) 
\chi_{n}^{(+)}\\
+\sum _{n=1}\beta _{n}(q)\chi_{n}^{(-)}~~,
\label{chi12}
\end{equation}
where spinors $\chi_n^{(\pm)}$ are equal to
\begin{equation}
\chi_n^{(+)} =
\begin{pmatrix}
\phantom{\frac{\displaystyle g}{\displaystyle L'g}}ce_{n}(\phi) Le_n(z,q)\\
-\frac{\displaystyle i}{\displaystyle E}\frac{\displaystyle
ce_{n}(\phi)Le_n'(z)+i\,ce'_{n}(\phi)Le_n(z)}{\displaystyle a\sinh z\cos \phi - a\cosh z\sin\phi}
\end{pmatrix}~.
\label{chi1}
\end{equation}
and
\begin{equation}
\chi_n^{(-)} =
\begin{pmatrix}
\phantom{\frac{\displaystyle g}{\displaystyle L'g}}se_{n}(\phi) Lo_n(z,q)\\
-\frac{\displaystyle i}{\displaystyle E}\frac{\displaystyle
se_{n}(\phi)Lo_n'(z)+i\,se'_{n}(\phi)Lo_n(z)}{\displaystyle a\sinh z\cos \phi - a\cosh z\sin\phi}
\end{pmatrix}.
\label{chi2}
\end{equation}
So far we almost literally followed the route of the single half-flux
problem. However, even a cursory  examination of the solutions $\chi^{(\pm)}_n$ shows
that something is amiss. In obtaining (\ref{gensolutionfluxes}) and then 
$\chi^{(\pm)}_n$, we never discarded any solutions, and yet the upper
components of all $\chi_n^{(\pm)}$ are always perfectly regular. So, where are the
wavefunctions with upper components divergent near the half-fluxes? 
 
Before we answer this question, let us examine the
lower components of the solutions. The elliptical 
coordinates of the flux located at $(x,y) = (a,0)$ 
are $(z,\phi) = (0,0)$. Therefore, in the vicinity of this flux, we have
\begin{equation}
\frac1{ a\sinh z\cos \phi - a\cosh z\sin\phi} \approx \frac1{a (z-i\phi)}
\approx \frac{e^{i\theta_1/2}}{\sqrt{2a\rho_1}},
\label{denom1}
\end{equation}
where $\rho_1$ is the distance between the point $(x,y)$ and the flux, and
$\theta_1\in (-\pi,\pi)$ is the polar angle shown in Fig.~\ref{Fig:intproblem}.
Similarly, near the second flux at $(-a,0)$, we have $(z,\phi)\approx (0,
\pi)$, and therefore
$$
\frac1{ a\sinh z\cos \phi - a\cosh z\sin\phi} \approx \frac1{a (z-i\phi)}
\approx \frac{e^{i\theta_2/2-i\pi/2}}{\sqrt{2a\rho_2}} .
$$
Since $Le_n(z=0,q)=0$ and $Lo_n'(z=0,q)=0$, in the vicinity of the first
flux, the divergent part of $\chi_n^{(\pm)}$ is 
\begin{align}
\chi_n^{(+)} &=
-\frac{\displaystyle i\,ce_{n}(0,q)Le_n'(0,q)}{\displaystyle E} \frac{e^{i\theta_1/2}}{\sqrt{2a\rho_1}}
\begin{pmatrix}
0\\
1
\end{pmatrix}~~,\\
\chi_n^{(-)} &=
\frac{\displaystyle se'_{n}(0,q)Lo_n(0,q)}{\displaystyle E} \frac{e^{i\theta_1/2}}{\sqrt{2a\rho_1}}
\begin{pmatrix}
0\\
1
\end{pmatrix}~~.
\label{divpart1}
\end{align}

The divergent part of the wavefunctions near the second flux is almost
identical:
\begin{align}
\chi_n^{(+)} &=
-\frac{\displaystyle i\,ce_{n}(\pi,q)Le_n'(0,q)}{\displaystyle E}
\frac{e^{i\theta_2/2-i\pi/2}}{\sqrt{2a\rho_2}}
\begin{pmatrix}
0\\
1
\end{pmatrix}~~,\\
\chi_n^{(-)} &=
\frac{\displaystyle se'_{n}(\pi,q)Lo_n(0,q)}{\displaystyle E}
\frac{e^{i\theta_2/2-i\pi/2}}{\sqrt{2a\rho_2}}
\begin{pmatrix}
0\\
1
\end{pmatrix}~~.
\label{divpart2}
\end{align}
Comparison with the single half-flux boundary condition at the flux location
(\ref{divergent1flux}) suggests that the wavefunctions $\chi^{\pm}_n$, which
we just constructed, are merely one of many possible self-adjoint extensions.
This extension contains wavefunctions which near both fluxes have regular
upper components and divergent lower component. 

What about the other self-adjoint extensions?
Where did we lose them? After all, we only 
repeated the steps for the single flux
Aharonov-Bohm problem, where this approach allowed us to find all
self-adjoint extensions.  Why didn't they naturally appear from
(\ref{constructEF})? 

To understand what went wrong, consider again solutions of the Schr\"odinger
equations for the single half-flux (\ref{genJ}) and compare them to general
solution of the Schr\"odinger solution for the two half-fluxes problem
(\ref{gensolutionfluxes}). One glaring distinction is that the former
contains solutions divergent near the fluxes, while the latter does not! To be
sure, there are certainly solutions of the two-flux problem that diverge
near the fluxes -- moreover, we constructed them explicitly, when we found the
lower components of $\chi^{\pm}_n$. The paradox appeared because
these divergent solutions do {\em not} have a separable form
$f_1(z)f_2(\phi)$ because the fluxes are point-like defects.  The line
connecting the two fluxes, and the elliptical coordinates used to describe
it were just tools, which allowed us to analyze the problem analytically,
but the underlying structure  of the divergencies is still that of point
like defects. They are, naturally,  most easily described in local polar
coordinate system around each flux. The functional form of the 
singularities, which has a structure
$F_1(\rho_i)F_2(\theta_i)$, is incompatible with a factorization
$f_1(z)f_2(\phi)$, as can be seen for example from (\ref{denom1}).

In short, we were able to find the self-adjoint extension Eq.
(\ref{chi12})-(\ref{chi2}) analytically in elliptical coordinates
so easily only because we started with the regular upper components.
Then it was easy to construct a complete basis of wavefunctions
of the form $ce_n(\phi)Le_n(z)$ and $se_n(\phi)Lo_n(z)$. Only then we
worked out the lower components, which ended up divergent near flux
locations, but not factorizable.

Can we find other self-adjoint extensions using elliptical
coordinates? There is one other case that is easy to solve:
wavefunctions with the regular lower component. Rather than using
(\ref{constructEF}), we could have tried the form
\begin{equation}
f=
\begin{pmatrix}
\frac{1}{E} (p_x-ip_y)v(x,y)\\
v(x,y)
\end{pmatrix}~~.
\label{constructEF2}
\end{equation}
We would start from the regular solutions of the wave equation
for $v$:
$$
v(\phi,z) = \sum \alpha_n ce_n(\phi) Le_n(z)+ \beta_n se_n(\phi) Lo_n(z)~~,
$$
and then would apply operator $(p_x-ip_y)/E$ to find the upper
components of the spinors forming the basis. This naturally would
result in a self-adjoint extension with the regular lower
components, and divergent upper components -- the opposite of the
first self-adjoint extension we found in (\ref{chi12}-\ref{chi2}). 
Note that these two self-adjoint extensions just found,  
with the regular upper or
lower components, are the ones that follow from the simplest physical
regularizations \cite{alford} of the problem obtained 
by replacing each of the  the infinitely
thin Aharonov-Bohm strings by a solenoid of finite radius.

Can one do better and construct the basis analytically 
for the general case, with
arbitrary parameters $\theta_{1,2}$ characterizing via
(\ref{divergent1flux}) each of the fluxes? At present, we
do not know the answer to this question, and leave this 
interesting problem for future study.

\begin{thebibliography}{99}
\bibitem{footi} Of course, some of the low energy states are strongly peaked
near a vortex. Even such states, however, do ``leak out'' 
into the continuum and
develop power law decay, mostly along nodal directions.
\bibitem{footii} The reader will realize that
the linearized approximation discussed here
is valid only sufficiently far from vortex
cores. For many purposes, however, this is precisely what is needed since
the thermodynamics and quasiparticle 
transport at low fields ($l\gg\xi$) are dominated by properties
of nodal quasiparticles in the bulk, far away from vortex locations.
The effective theory corresponding to
this linearized version of the BdG problem generally exhibits
characteristic universal scaling of quasiparticle 
thermodynamics and transport properties, the dimensionless 
scaling parameter being the ratio of thermal to
magnetic length\cite{simonlee}. The actual form of various 
scaling functions and quasiparticle energy spectrum  
follow directly from the approach of FT\cite{ft00}.
\bibitem{footiii} Throughout the paper we assume 
that $k_F^{-1}$ is smaller than
but still roughly comparable to $\xi$ since this 
is the case for cuprate superconductors. Both 
lengths are far shorter than magnetic length $l$ in the low-field regime
discussed here ($l\gg\xi\agt k_F^{-1}$).
\bibitem{phiaphib} An explicit expression for $\phi_{A(B)}(\br)$ in terms
of the antisymmetric elliptic theta function 
can be found in Ref. \onlinecite{marinelli}.
\bibitem{footB}{Although formally all eigenfunctions 
of the spectrum diverge near vortices
according to (\ref{AsymptA}) or (\ref{AsymptB}), the prefactor of the
divergent part of the wavefunctions rapidly decreases as a function of the
energy $|E|$. Therefore in practice the singular $r^{-1/2}$
behavior of the wavefunctions describing higher energy bands can be detected
only at distance very near a vortex. For sufficiently high energies, such distance
becomes comparable to the vortex core size, where the linearization is not
valid.}
\bibitem{foot_symlines} The points of the BZ with 
crystal momentum belonging to the symmetry
lines $k_x=-k_x$ and $k_y=-k_y$ ($mod~2\pi/l$)  
but do not coincide with $X$, $\Gamma$, $M$
and $Y$ are invariant under $Cm_x$ ($Pm_y$), however it is easy to show that
the representations of this subgroup are one-dimensional and do not result
in degeneracy of the states with such momenta.
\bibitem{foot_ashvin} The proof of the degeneracy of the spectrum by
Vishwanath\cite{ashvini,ashvinii} assumes periodicity of quantity 
$\Phi^2_{\{q\}}$, denoted by $\exp(-\delta\phi)$ by this
author. Actually, this quantity is anti-periodic 
as shown in the Appendix. Nevertheless,
the conclusions regarding the 
degeneracy of the spectrum at points $\Gamma$
and $M$ remain valid as follows from (\ref{repeqs}).
\end{thebibliography}

\end{document}